\documentclass[useAMS,usenatbib]{mn2e}
\usepackage{graphicx}
\usepackage{natbib}

\title[Matter distribution in galaxy clusters]{The re-distribution of matter in the
cores of galaxy clusters}
\author[Laporte et al.]{
\parbox[t]{\textwidth}{
Chervin F. P. Laporte$^{1,2}$ \& Simon D.M. White$^{1}$}
\\
$^{1}$ Max Planck Institute for Astrophysics, Karl-Schwarzschild-Strasse 1, 85740 Garching, Germany\\
$^{2}$ Department of Astronomy, Columbia University, 550 West 120th Street, New York, NY, 10027, U.S.A\\
}
\begin{document}
\date{}
\pagerange{\pageref{firstpage}--\pageref{lastpage}} \pubyear{2011}
\maketitle
\label{firstpage}
\begin{abstract}

We present cosmological N-body resimulations of the assembly of the
Brightest Cluster Galaxies (BCGs) in rich clusters. At $z=2$ we
populate dark matter subhalos with self-gravitating stellar systems
whose abundance and structure match observed high-redshift galaxies.
By $z=0$, mergers have built much larger galaxies at cluster centre.
Their dark matter density profiles are shallower than in corresponding
dark-matter-only simulations, but their total mass density profiles
(stars + dark matter) are quite similar. Differences are found only at
radii where the effects of central black holes may be significant.
Dark matter density slopes shallower than $\gamma=1.0$ occur for
$r/r_{200} < 0.015$, close to the half-light radii of the BCGs. Our
experiments support earlier suggestions that NFW-like profiles are an
attractor for the hierarchical growth of structure in collisionless
systems -- total mass density profiles asymptote to the solution found
in dark-matter-only simulations over the radial range where mergers
produce significant mixing between stars and dark matter.  Simulated
dark matter fractions are substantially higher in BCGs than in field
ellipticals, reaching 80\% within the half-light radius. We also
estimate that supermassive black hole mergers should create BCG cores
as large as $r_{c}\sim 3 \,\mathrm{kpc}$.  The good agreement of all
these properties with recent observational studies of BCG structure
suggests that dissipational processes have not played a dominant role
in the assembly of the observed systems.

\end{abstract}

\section{Introduction}
In the standard $\Lambda$CDM cosmological paradigm, dark matter is a
collisionless particle and its clustering can be followed by solving
the collisionless Boltzmann equation (CBE). This can be done using
Monte-Carlo techniques such as the N-body method. In the last decade,
cosmological dark-matter-only simulations have shown that the
spherically averaged density profiles of dark matter haloes follow a
universal form from clusters to dwarf galaxy scales which may be
parametrised by the \citet*['NFW' hereafter]{Navarro1996, Navarro1997}
profile\footnote{It should be noted however that some scatter exists
  from halo to halo \citep{Navarro2010} and the asymptotic slope may
  be go to lower values but at radii much smaller than the typical
  half-light radius of galaxies, by an order of 100.}. This density
profile falls off as $\rho\propto r^{-3}$ at large radii and
asymptotes to $\rho \propto r^{-1}$ as $r\to 0$. The origin of this
regularity is still unclear. Many arguments have been put forward,
ranging from the effects of from multiple mergers\citep{Syer1998} to
the maximisation of entropy subject to constraints like constancy of
the actions \citep{Pontzen2013} but a convincing demonstration of the
emergence of a universal profile is still lacking.

Observations of relaxed galaxy clusters, on the other hand, suggest
that the dark matter density profiles in the innermost regions are
shallower than those found in simulations \citep{Sand2002, Sand2004,
  Sand2008, Newman2009, Newman2011, Newman2013b}. Intriguingly, recent
results from \cite{Newman2013a} combine a variety of mass measurement
techniques (stellar kinematics, strong lensing, weak lensing and X-ray
observations of the hot intracluster gas) to show that the {\it total}
mass density profile in these regions (dark matter + stars + gas) does
appear to be consistent with that found in dark-matter-only
simulations. This has been interpreted as showing that although
dissipational galaxy formation physics may compress or expand dark
matter haloes during the early condensation and star formation phases,
later collisionless merging re-establishes the NFW profile, which can
thus be thought of as an attractor for the violent relaxation of
collisionless systems. This hypothesis for the assembly of Brightest
Cluster Galaxies (BCGs) was originally suggested by \cite{Loeb2003}.
Although baryon condensation into stars steepens the inner regions of
the total density profiles of the progenitors, the many
dissipationless mergers required to form the BCG take the combined
collisionless fluid back to the universal profile found in dark-matter
only simulations. \cite{Gao2004} explored this process qualitatively
by tracking the central dark matter particles in dark matter-only
simulations of galaxy clusters back to higher redshifts, demonstrating
that the even the innermost dark matter distribution is indeed built
up by the merging of a large number of separate entities present at
higher redshift. They suggested that, in the presence of stars,
efficient mixing would reestablish an NFW-like profile. However, an
experiment containing two self-gravitating collisionless fluids (dark
matter and stars) with initial conditions constrained by observations
of real high-redshift galaxies needs to be carried out to test whether
the attractor hypothesis is effective in a realistic situation.

It is generally thought that as dynamical friction brings the visible
components of galaxies to cluster centre, it will evacuate dark matter
from this region, pushing it to larger radii
\citet[e.g.][]{El-Zant2004}. However, dynamical friction and tidal
stripping affect both the stars and the dark matter, both in the
central object and in the galaxy which is merging with it.
\cite{Laporte2012} showed that the competition between these processes
can lead to different outcomes depending on the initial structure of
high-redshift galaxies. In the simulations they analysed, these
authors found that the mixing of stars and dark matter through
dissipationless mergers did indeed reduce the slope of the dark matter
cusp by up to $\Delta\gamma=0.5$ at the innermost resolvable radius,
but their initial conditions were inconsistent with modern
observations of high-redshift galaxies, so their results could only be
interpreted qualitatively.

In this paper, we test whether the growth of galaxy clusters in a
$\Lambda$CDM universe, starting from a $z=2$ galaxy population
consistent with that observed, can produce realistic low-redshift
BCGs, in particular, with dark matter cusps as shallow as those of
\cite{Newman2013b} and total mass density and luminosity profiles
consistent with \citep{Newman2013a}. We run simulations which
explicitly emulate the compression of halo structure at $z=2$ caused
by galaxy condensation, and we follow the subsequent evolution of the
star and dark matter distributions fully self-consistently (under the
hypothesis of no further star formation) to the present-day. To this
end, we have developed a method for inserting self-consistent stellar
components into dark matter haloes formed in a cosmological
simulation. Past simulations of this sort have replaced dark matter
haloes (and all their substructure) with spherical compound galaxy
models made up of stars and dark matter \citep{Dubinski1998,
  Rudick2006, Ruszkowski2009}.  We start with the $z=2$ output of a
dark-matter only zoom-in simulation and insert equilibrium stellar
spheroids directly into the centres of the dark matter
subhlaos. Galaxy properties are as inferred from abundance matching
results \cite{Moster2013} and from observed mass-size relations at
$z=2$. The systems re-equilibrate on a short timescale as the dark
matter adjusts to the potential of the newly inserted stars, and
thereafter the full system can be followed self-consistently to $z=0$
and the results compared to those of the original dark-matter-only
simulation.

Section 2 introduces the original cluster simulations, discusses our
method for inserting equilibrium stellar components into their $z=2$
subhaloes, and tests that the results of this procedure are indeed
stable, in the absence of merging, over cosmological timescales. We
study the structural properties of our BCGs/clusters at $z=0$ in
section 3. In section 4 we look more closely at the mergers and mixing
processes that occur during the assembly of the central regions of the
clusters, and we discuss the important role of black holes in section
5. Sections 6 and 7 provide further discussion and set out our
conclusions, respectively.

\section{Numerical methods}
\subsection{Simulations}

We use a suite of zoom-in dark-matter-only simulations of galaxy
clusters, the {\sc phoenix} project \cite{Gao2012b} as our starting
point for re-simulating the passive evolution of galaxies from $z=2$
to $z=0$. The haloes in the {\sc phoenix} suite were initially
selected from the {\it Millennium Simulation} \citep{Springel2005b}
and were re-simulated at a variety of resolutions. In this paper we
use resolution level 2 simulations of two of the nine Phoenix clusters
(simulations Ph-C-2 and Ph-E-2 in the original notation).  These
have comoving softening length $\epsilon=0.3 \,h^{-1}\, \mathrm{kpc}$
and mass resolution $m_{p}\sim4-10\times10^{6} \,h^{-1} \, {\rm
  M_{\odot}}$. Further details of the simulations are given in
\cite{Gao2012b}. They adopt the cosmology of the original Millennium
Simulation: $\Omega_{m}=0.25$, $\Omega_{\Lambda}=0.75$,
$\sigma_{8}=0.9$ and $n=1$. These parameters are outside the range
currently considered plausible, but this is of no conseqence for the
current paper. For the rest of the discussion, our spatial and mass
units are in $\rm kpc$ and in $\rm M_{\odot}$ respectively (i.e. no
$h$). Subhaloes were identified using the {\sc Subfind}
algorithm\citep{Springel2001}. We insert equilibrium models of the
stellar components of galaxies into the dark matter subhaloes of the
original simulations at $z=2$, and follow the later evolution of the
stellar and dark matter distributions down to the present-day. For
these resimulations we keep the softening lengths for star and dark
matter particles fixed in physical units at $\epsilon=0.1 \,
\mathrm{kpc}$, corresponding to the softening of the original dark
matter simulation at $z=2$.

\subsection{Generating the stellar components of pre-existing halos}

The galaxies are represented by \cite{Hernquist1990} spheres:
\begin{equation}
\rho_{*}=\frac{aM_{*}}{r\left(r+a\right)^3},
\end{equation}
where a is the scale radius which is related to the 3D half-mass
radius through $a=r_{e}/(\sqrt{2}+1)$. The half-mass radius in
projection is related to $r_e$ through $R_{e}=r_{e}/1.33.$

Initially spherical N-body models for the stellar distributions are
generated through Monte Carlo sampling of an isotropic distribution
function (DF) of the form $f\equiv f(E)$ using a von Neumann rejection
technique \citep{Kuijken1994,Kazantzidis2004}. The DF takes the form:
\begin{equation}
f_{*}(\mathcal{E})=\frac{1}{\sqrt{8}\pi^2}
\int^{\mathcal{E}}_{0}\frac{{\rm
d}\Psi}{\sqrt{\mathcal{E}-\Psi}}\frac{{\rm d}^2\rho_{*}(\Psi)}{{\rm
d}\Psi^2} + \frac{1}{\sqrt{\mathcal{E}}}\frac{{\rm d}\rho_{*}}{{\rm
d}\Psi}\bigg|_{\Psi=0},
\end{equation}
where $\Psi=-\Phi+\Phi_{0}$ and $\mathcal{E}=-E+\Phi_{0}=\Psi-v^2/2$
are the potential relative to halo centre (containing the
contributions of both dark matter and stars) and the specific orbital
energy, respectively. The potential of the dark matter halo is
modelled as that of the Hernquist sphere which best fits (in a
$\chi^2$ sense) the spherically averaged density profile of the
halo. Because the total potential is modelled as the superposition of
two Hernquist spheres, the second term in equation 2 drops out. While
there are better ways to represent the potential of the dark matter
haloes, we have found this approximation to be robust enough to
produce stable galaxies (as will be shown later). Moreover, we note
that although it is often recommended to take account of the Plummer
softening when generating N-body models from distribution function
based methods, in the case of a Hernquist potential, \cite{Barnes2012}
demonstrates that this is not necessary provided the softening length
is smaller than the scale radius of the galaxy and that the stellar
profile is not strongly centrally cusped (e.g. $\rho_\star\propto
r^{-2}$). The spherical equilibrium stellar distribution produced by
this procedure is centred on the potential minimum of simulated halo
and given a mean velocity equal to that of the halo's inner regions.

\subsection{Stability tests}

We perform some stability tests to check whether our method works
properly. We extract a subhalo identified by {\sc subfind} at $z=2$ in
the cosmological dark-matter-only run of Ph-E-2, and we insert a
spheroid with stellar mass and size determined as described in the
next subsection. We then evolve this system in isolation with a fixed
physical softening length of $\epsilon=0.1 \, {\rm kpc}$ for a period
of 10 Gyr. The dark matter contracts on a timescale of 200 Myr due to
the gravitational effects of the impulsively imposed stellar
component, but the stellar component itself varies much less because
its own potential was included when initialising particle
velocities. This is shown in Figure 1 which shows that even after 10
Gyr the half-light radius of the galaxy has only changed by $\sim
10\%$ and its luminosity profile is still quite similar to that set up
initially.  The initial change of the dark profile can be seen in
Figure 2. There is a substantial steepening throughout the region
occupied by stars, but this compressed profile is then stable over Gyr
periods. The nonspherical shape of the dark mater halo also induces a
flattening of the stellar component after re-virialisation, as shown
in Figure 3. We have not attempted to prevent the rapid contraction of
the dark matter component, considering that the ``adiabatically
compressed'' dark matter profile of the later panels of Figure 2 is as
likely to be realistic as any alternative which we could set up. Note
that this initial contraction implies that the dark matter
distributions of our galaxies have significantly {\it steeper} inner
profiles (and the total mass distributions substantially steeper inner
profiles) than the corresponding objects in the dark-matter-only
simulation. This makes all the more remarkable the relaxation towards
NFW total mass profiles which we demonstrate below.

\begin{figure}
\includegraphics[width=0.5\textwidth,trim=0mm 0mm 0mm 0mm,clip]{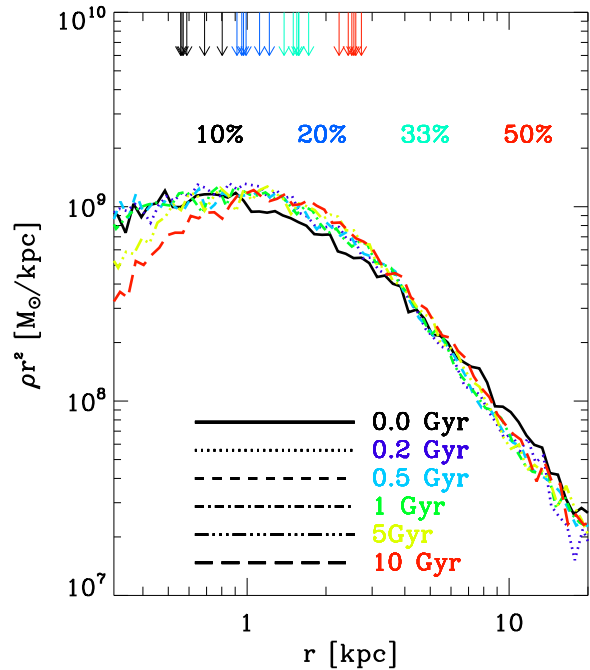}
\caption[Stability test: stellar density profile evolution]{Evolution
of the stellar density profile in our stability test. Using the
machinery described in the text, we introduce a stellar component in a
live dark matter halo which we evolve in isolation. The system
undergoes a rapid phase of re-virialisation which only slightly
changes the stellar density profile. Thereafter it changes very little
over 10 Gyr of evolution. Arrows mark the radii enclosing 10, 20 33
and 50 percent of the light, these increase with time, but the overall
increase in half-light radius is of order 10 percent. This galaxy has
a dark halo mass of $M_{h}=7\times10^{12} \, \mathrm{M_{\odot}}$, a
stellar mass of $M_{*}=9\times10^{10}\, \mathrm{M_{\odot}}$ and a
half-light radius of $r_{e}\sim 2.5 \,\mathrm{kpc}$}
\end{figure}

\begin{figure}
\includegraphics[width=0.5\textwidth,trim=0mm 0mm 0mm
  0mm,clip]{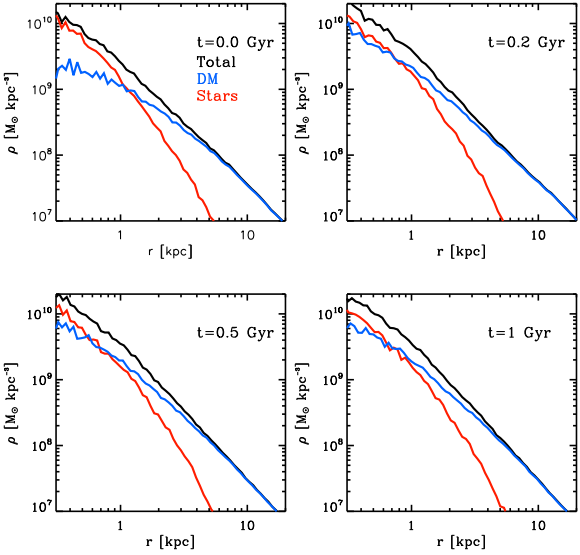}
\caption[Stability test: stellar, dark and total density
  profile]{Evolution of the density profiles of the stars (red), the dark
  matter (blue) and the total mass (black) in our stability test.
  The system stabilises quickly after an initial contraction of the
  dark matter on a timescale of $t\sim 200$ Myr.}
\end{figure}

This method improves significantly on past attempts to embed galaxies
in dark-matter-only cosmological simulations. These typically replaced
entire simulated dark matter haloes by idealised equilibrium models of
a galaxy embedded in a smooth and spherical dark matter halo
\citet{Dubinski1998, Rudick2006, Ruszkowski2009}, thus losing the true
shape and the substructure of the original halo and causing artifacts
at the boundaries between embedded haloes and their surroundings. In
our experiment, the original halos are retained and settle into new
quasi-equilibrium configurations after a few central dynamical
times. In addition, recent improvements in computer capabilites mean
that we are able to carry out experiments of significantly higher
resolution than was previously possible, allowing us to follow the
structure of galaxy/halo systems realistically down to subkiloparsec
scales. Based on a series of tests similar to that illustrated here
but for galaxies of different stellar mass, we conclude that we can
reliably follow the structural evolution of galaxies with $M_{*}
\geq 2 \times 10^{10} {\rm M_{\odot}}$ without the effects of
softening or two-body relaxation compromising our results.

\begin{figure}
\includegraphics[width=0.5\textwidth,trim=0mm 0mm 0mm 0mm,clip]{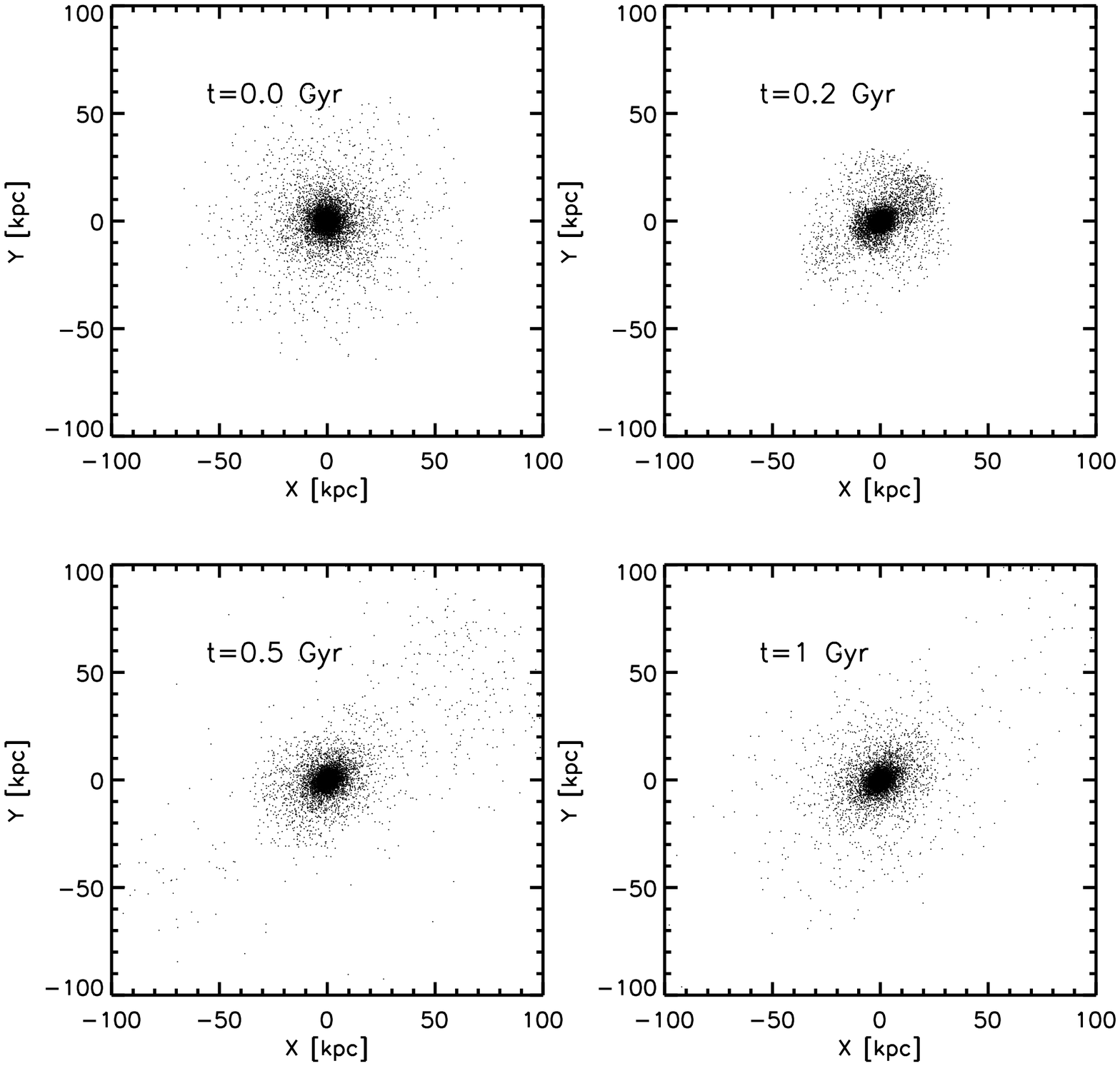}
\caption[Stability test: x-y projection of stars]{Projected images of
  the stellar component of the same galaxy as in Figure 1 and 2,
  showing the flattening induced by the re-virialisation process. The
  stellar distribution is centred on the potential minimum of the dark
  matter halo, and the mean velocity of the 100 most bound dark matter
  particles is added to that of each of the stars to ensure that
  galaxy and halo are initially moving together. The image illustrates
  that the time-dependent inner structure of real dark matter haloes
  can drive some initial transients in the stellar distribution, but
  these lead to relatively minor long-term effects.}
\end{figure}

\subsection{Initial conditions}                                                        
Using the method described above, we put galaxies at the centres of
all $z=2$ subhaloes in the original dark-matter-only simulations which
are expected to host galaxies with $M_\star >
2\times10^{9}\,\mathrm{M_{\odot}}$. We wish to insert a galaxy
population that agrees closely with that observed at $z=2$. The
stellar mass of each galaxy is therefore chosen so that it lies on the
$M_{200}-M_{*}$ relation inferred from the abundance-matching analysis
of \cite{Moster2013}. Its size is then set according either to the
mass-size relation for red, compact, massive and quiescent galaxies or
to that of extended, blue, star-forming galaxies as given by
\cite{vanderWel2014}. The choice between the red and blue relations is
made in a probabilitistic way using the observed fraction of quiescent
$z=2$ galaxies as a function of stellar mass from \cite{Muzzin2013}
which is determined down to $10^{10} \, \mathrm{M_{\odot}}$. Below
this stellar mass, sizes are selected according to the overall
scatter in the mass-size diagram of galaxies (which is
substantial). The result is illustrated in Figure 4. Although 
it is clearly still an approximation to represent all galaxies as Hernquist
spheres, our experiment is still realistic in that it assigns
every dark matter subhalo a galaxy with a stellar mass and size which
are consistent with observations of $z=2$ galaxies.

\begin{figure}
\includegraphics[width=0.5\textwidth,trim=0mm 0mm 0mm 0mm,clip]{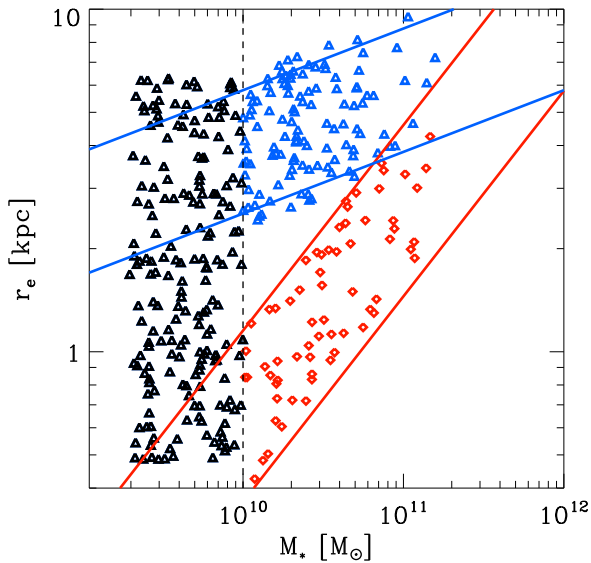}
\caption[The $z=2$ mass-size relation]{Mass-size relation of the
  galaxies we insert into the Ph-E-2 simulation. Red symbols have
  properties set according to the observed relation for compact,
  massive, quiescent galaxies, while blue ones follow the observed
  relation for the massive, star-forming galaxies. Below $10^{10} \,
  \mathrm{M_{\odot}}$, observational classifications are uncertain so
  we mimic the overall observed scatter in size (black triangles).}
\end{figure}

\subsection{Cluster simulations}
We have resimulated two cluster mass haloes from the Phoenix
suite. These are the Ph-E-2 and Ph-C-2: their properties are
summarized in Table 1. Before starting a detailed presentation of our
results, it useful to understand where the material which makes up the
central $r<10 \mathrm{kpc}$ of the final system actually comes from.
This is shown in Figure 5, which illustrates graphically that the
inner regions of BCGs are expected to be built up from a significant
number of disjoint obects, as already pointed out using similar plots
by \cite{Gao2004}. In Figure 11 below we will show that this is true 
for the stars as well as for the dark matter.

\begin{table}
 \centering
 \begin{minipage}{130mm}
  \begin{tabular}{@{}llrrrrlrlr@{}}
  \hline
Run & $ M_{200}$ & c & $R_{200}$ & \\
 & $h^{-1} \rm{M_{\odot}}$ &  & $h^{-1} \mathrm{Mpc}$ \\
  \hline
Ph-C-2 & $5.495 \times 10^{14}$  & 5.11 & $1.386 $ \\
Ph-E-2 & $5.969 \times 10^{14}$  & 5.19 & $1.369 $ \\
\hline
\end{tabular}
\end{minipage}
\caption[Phoenix Clusters Properties]{Basic structural properties
  (virial mass $M_{200}$, concentration $c$ and virial radius
  $R_{200}$) of the Phoenix clusters at $z=0$ from the original
  dark-matter-only simulations.}
\end{table}

\begin{figure*}
\includegraphics[width=0.4\textwidth,trim=0mm 0mm 0mm 0mm,clip]{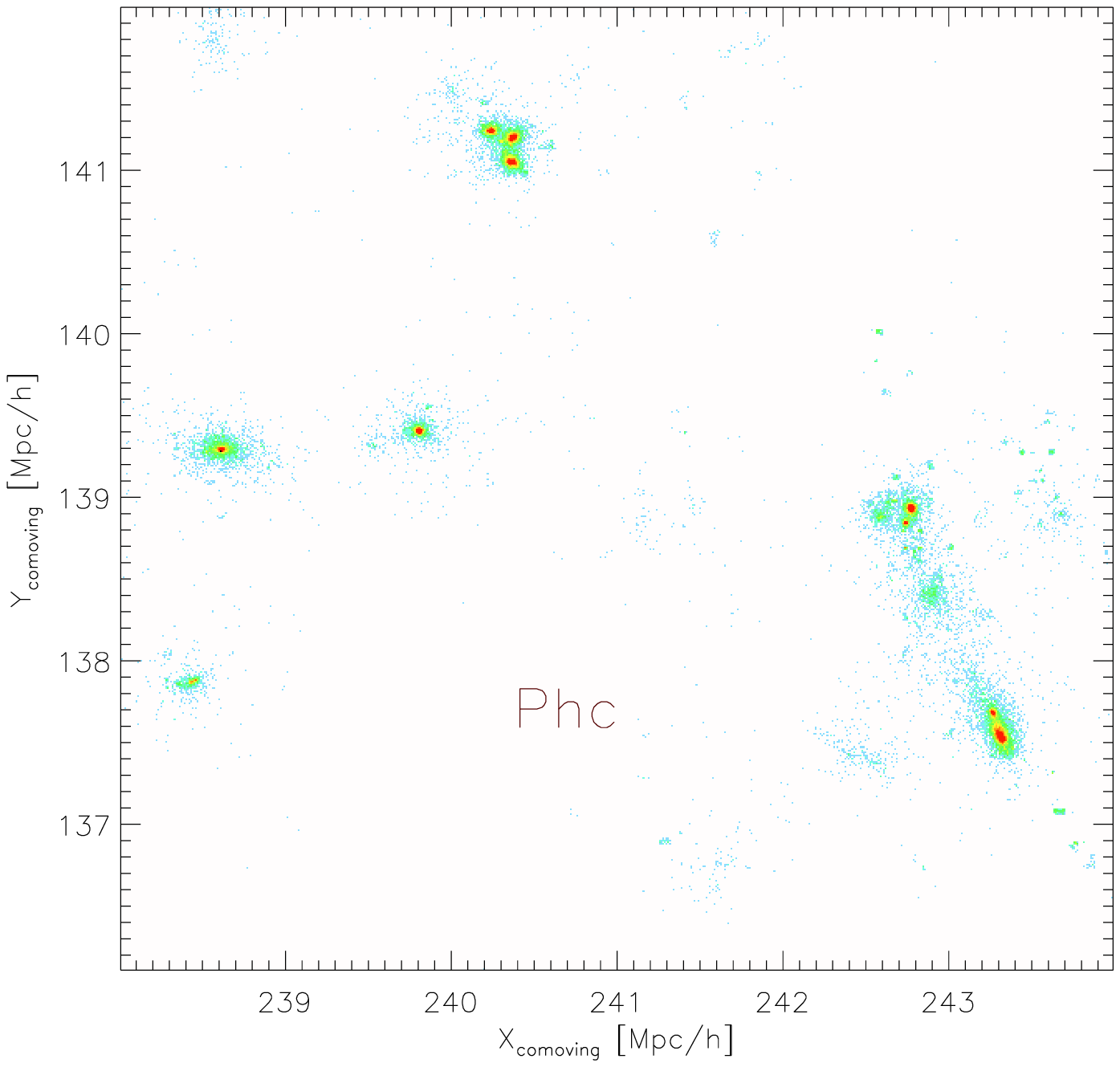}
\includegraphics[width=0.4\textwidth,trim=0mm 0mm 0mm 0mm,clip]{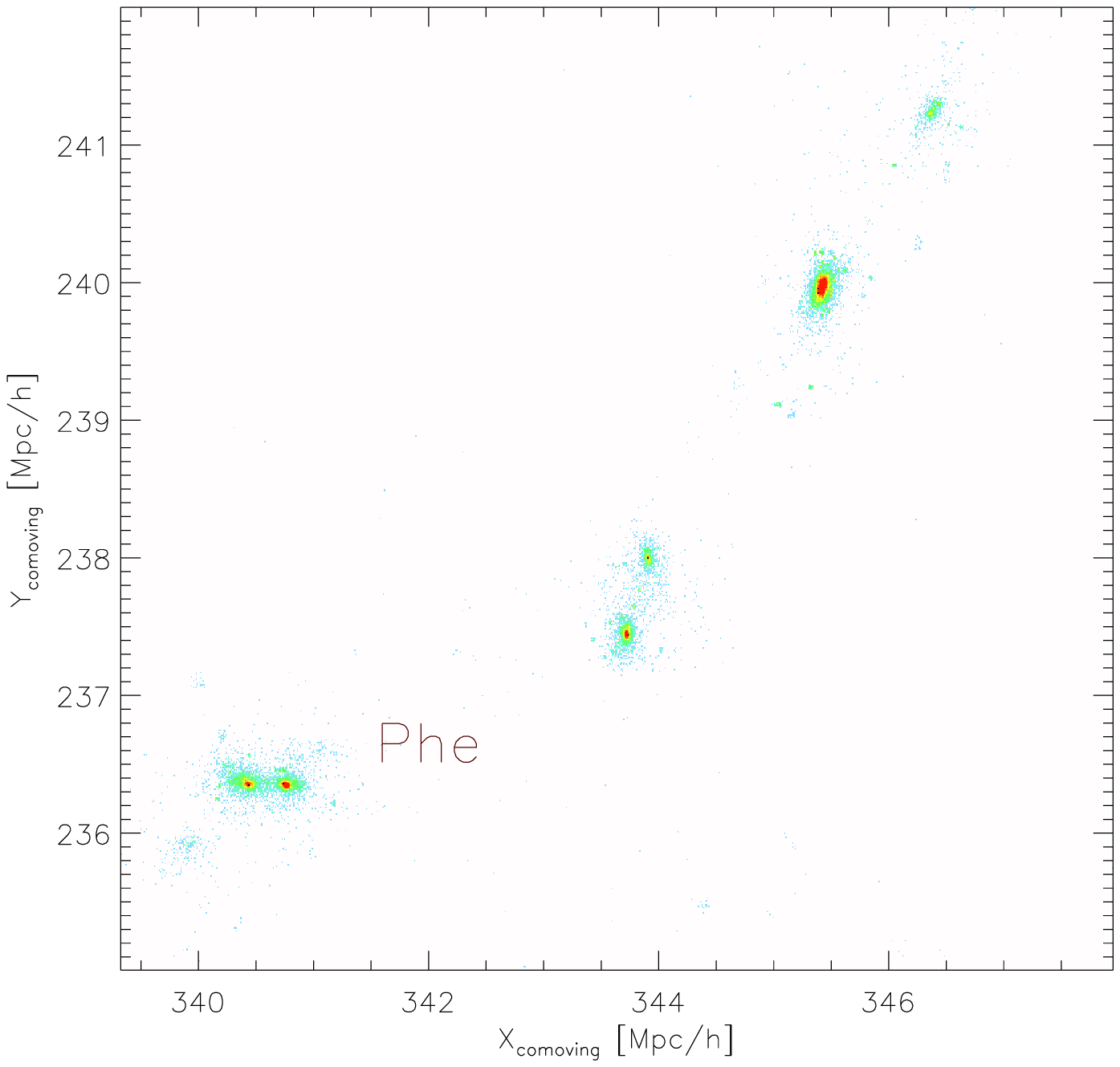}
\caption[Distribution of central $z=0$ 10 kpc DM particles at
  $z=2$]{Distribution at $z=2$ of the dark matter particles which at
  $z=0$ lie within 10kpc of cluster centre in the original {\it
    Phoenix} simulations Ph-C-2 (left) and Ph-E-2 (right). This
  illustrates that even the very central regions of galaxy clusters
  are typically in many disjoint pieces at $z=2$.}
\end{figure*}

\section{Structure of galaxy clusters}

\subsection{Density profiles}

In this section we present $z=0$ inner density profiles for the
stellar and dark matter components of our two simulations and for
their sum. We compare these to the profiles found in the corresponding
dark-matter-only simulations. These are plotted in Figure 6 in the
form of normalised $\rho r^{2}/(\rho_{200}r^{2}_{200})$ profiles as a
function of normalised radius $r/r_{200}$. Overplotted on the
simulations in this figure are the profiles derived by
\cite{Newman2013a,Newman2013b} for their sample of real clusters. The
first ineresting result to note is that the total density profiles in
the dark-matter-only and star+dark matter runs look very similar
despite the fact that they started out very different (compare the
blue and black profiles in the top left panel of Figure 2). The only
clear difference at $z=0$ is in the innermost regions of the
BCG. Furthermore, although the dark matter distributions after
revirialisation are clearly more centrally concentrated at high
redshift in the star+galaxy run than in its dark-matter-only
counterpart (compare the blue curve in the top left panel of Figure 2
with those in the other three panels) the opposite is true in the
final object, where the dark matter profile in the simulation with
stars dips below that in the simulation without stars for radii $r <
r_{200}\sim 0.01-0.02$, approximately the 3D half-light radius of the
final BCG ($r_{e}\sim 30 \, \rm{kpc}$ for Ph-C-2 and $r_{e}\sim 18 \,
\rm{kpc}$ for Ph-E-2, as shown by the red arrows). The stellar density
profile is also in reasonable agreement with the observed stellar
profiles of \cite{Newman2013a}. Since our simulations aasume purely
dissipationless evolution since$z=2$, this suggests that star
formation at lower redshifts may not play an important role in the
structuring of BCGs.

\begin{figure*}
\includegraphics[width=1.0\textwidth,trim=0mm 0mm 0mm 0mm,clip]{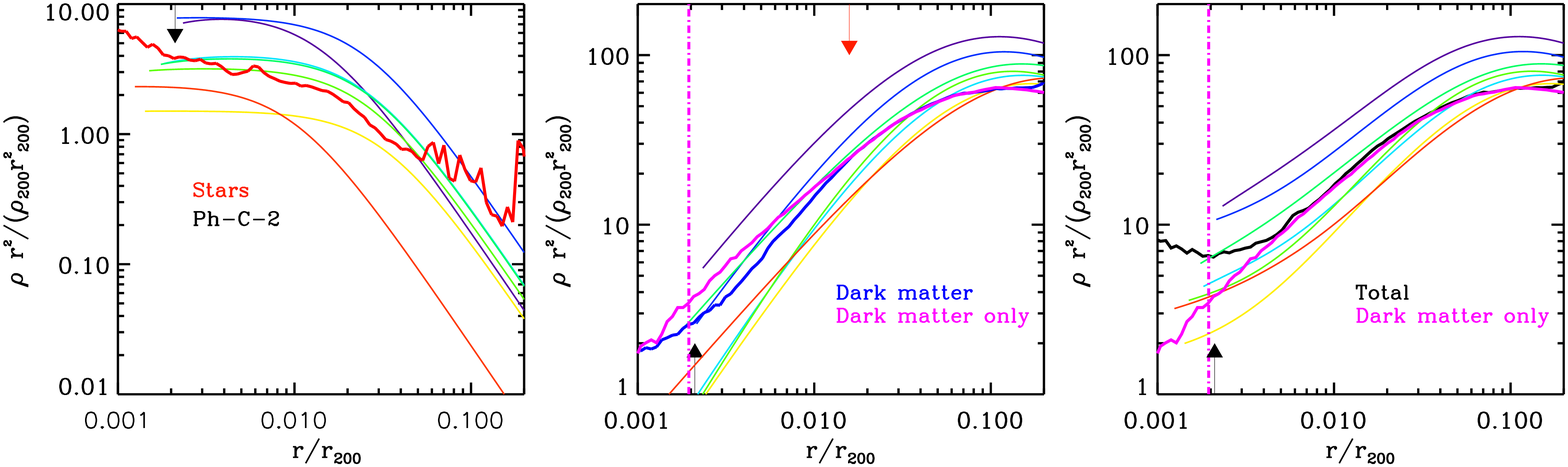}
\includegraphics[width=1.0\textwidth,trim=0mm 0mm 0mm 0mm,clip]{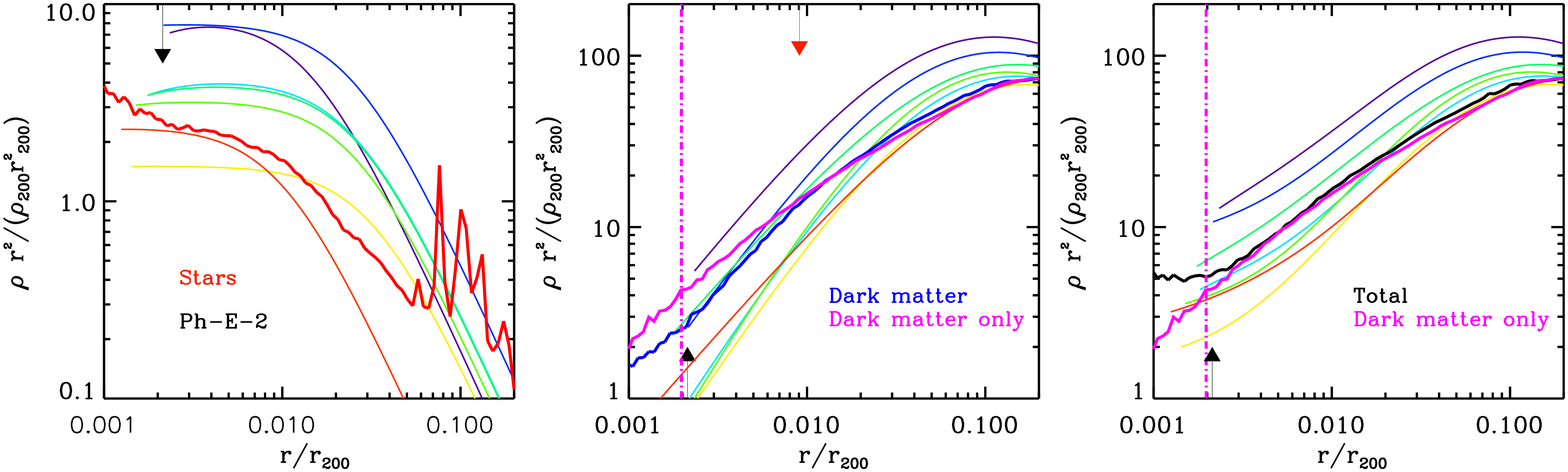}
\caption[Final density profiles for stars, dark matter and total
  matter.]{Heavy lines in each panel show $\rho r^{2}$ profiles for
  the clusters normalised by $\rho_{200} r_{200}^{2}$ as a function of
  normalised radius $r/r_{200}$ for the stars (left panel in red), for
  the dark matter (middle panel in blue) and for their sum (right
  panel in black). Thinner coloured lines show profiles for observed
  clusters from Newman et al. (2013a,b). The magenta lines in the
  middle and right panels show the profiles found in the original
  dark-matter-only simulations. The inclusion of stars has clearly
  caused a drop in the dark matter density at radii within $r\sim 10
  \, \mathrm{kpc}$. In contrast, the total mass profiles are very
  similar in the two cases, even well inside the half-light radii of
  the galaxies. Diiferences are only substantial in the innermost
  regions where the stellar density exceeds that of the dark
  matter. Black arrows mark the radii where a mass deficit (in the
  form of a core) would be expected due to black hole mergers (see
  section 6.4). Red arrows mark the half-light radii of the BCGs,
  which turn out to be close to the radii where the slope of the dark
  matter profile becomes shallower than $\gamma=1$.}
\end{figure*}

In Figure 7, we present density profile slopes, defined as
$\gamma=-\frac{\mathrm{d}\ln(\rho)}{\mathrm{d}\ln(r)}$ as a function
of normalised radius $r/r_{200}$. In computing the slopes we have
binned our data between $r=0.1\, \mathrm{kpc}$ and $r=1000 \,
\mathrm{kpc}$ into logarithmic bins of width $\Delta \log(r)=0.1$. We
have tried other binning schemes, finding they all lead to consistent
results. The slopes of our BCG stellar profiles are roughly consistent
with those measured, although they may be systematically to steep at
small radii $r/r_{200}<0.002$. The slopes of the dark matter density
profiles in our resimulations are systematically shallower than those
of the original dark-matter-only runs. The radius where these become
shallower than the asymptotic NFW value (represented by the horizontal
dash-dotted line) is $r\sim 0.01~r_{200}$ which agrees with that found
observationally ($0.01 <r/r_{200}< 0.03$ according to
\cite{Newman2013b}). This scale also corresponds closely to the BCG
half-light radius. Turning to the total density profiles, these also
match those observed by \cite{Newman2013a} except in the innermost
regions ($r< 0.002r_{200}$) where they clearly become steeper.  It
seems that mixing is efficiet enough to cause approach to an attractor
solution down to radii wwell within the half-light radii of the BCGs.

\begin{figure*}
\includegraphics[width=1.0\textwidth,trim=0mm 0mm 0mm 0mm,clip]{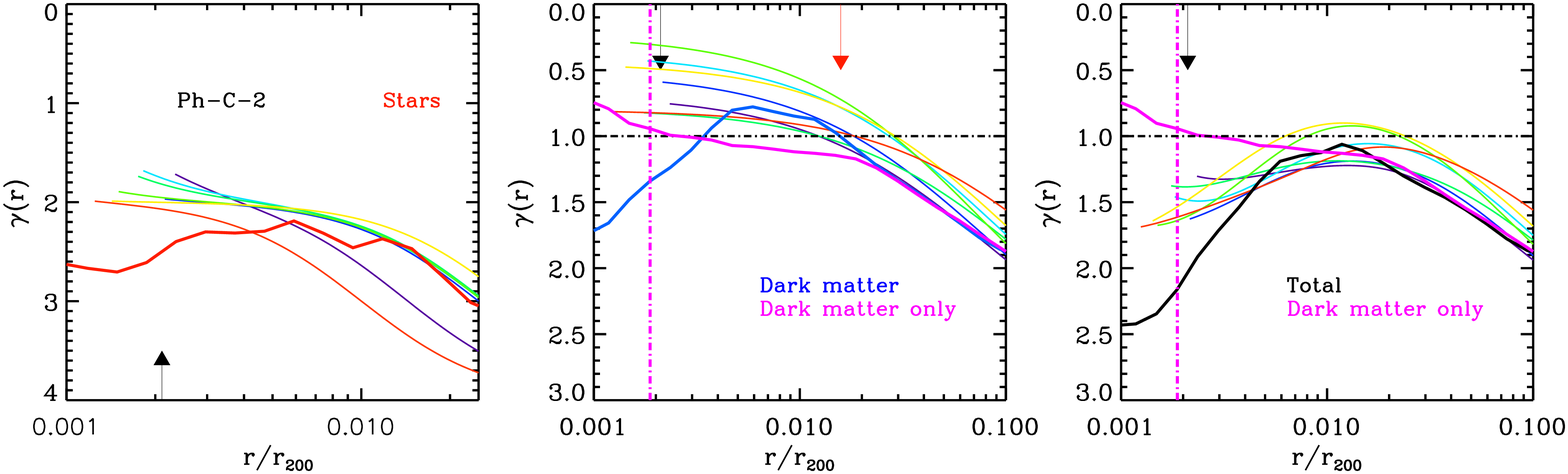}
\includegraphics[width=1.0\textwidth,trim=0mm 0mm 0mm 0mm,clip]{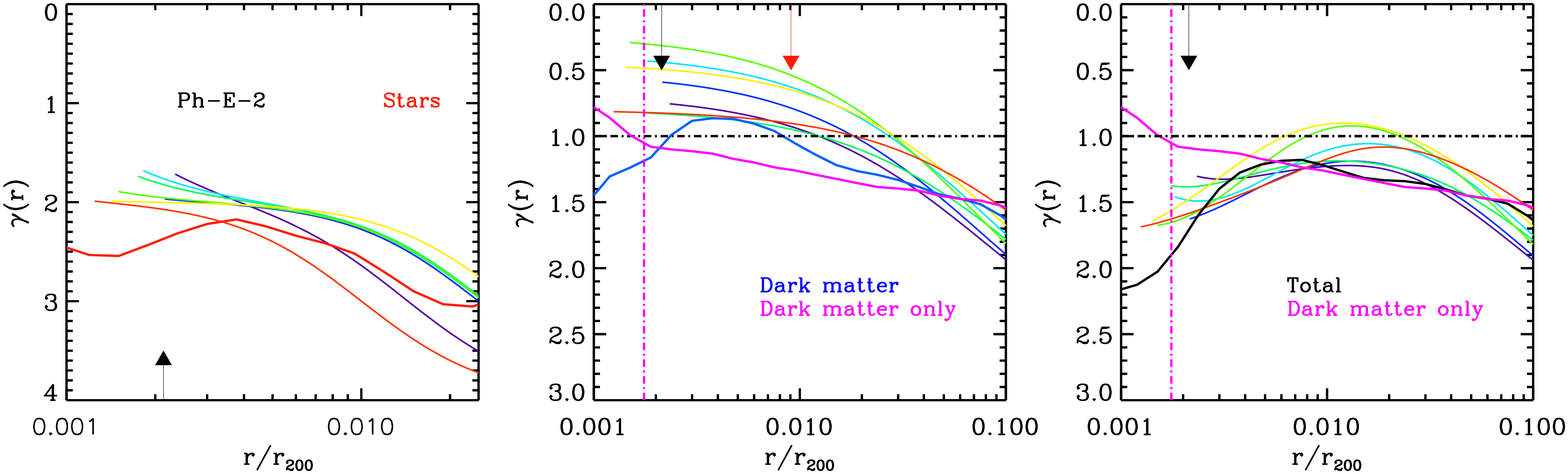} 
\caption[Slopes of the density profiles for stars, dark matter and
  total matter]{Density profile slopes as a function of normalised
  radius $r/r_{200}$ in the same format as Figure 6. Coloured lines
  are the slopes obtained for real clusters by Newman et
  al. (2013a,b). For the total mass, the agreement between observation
  and our resimulations is reasonably good (right panel). The dark
  matter slopes in our resimulations are clearly shallower than in the
  dark-matter-only runs (blue and magenta lines in the middle panel)
  and the scale at which they become less than $\gamma=1$ is similar
  to that found by Newman et al. (2013b) for real clusters,
  corresponding to physical radii of $r=18-30 \mathrm{kpc}$. Slopes
  for the stars are steeper and are also similar to those measured by
  Newman et al. (2013b).  Black arrows mark the radii where we
  estimate that a mass deficit (in the form of a core) due to black
  hole mergers should become appreciable (see section 6.4). Red arrows
  mark the half-light radii of the BCGs, which are similar to the
  radii within which $\gamma<1$ for the dark matter.}
\end{figure*}

\subsection{Velocity Dispersion Profiles}

In Figure 8, we compare projected stellar line-of-sight velocity
dispersion (LOSVD) profiles for our BCGs (based on 100 random
projections) to observations from \cite{Newman2013b}. We observe a
very similar rise in LOSVD with radius as reported by
\cite{Newman2013a} for their observed clusters, with central velocity
dispersion $\sigma\sim300 \, \rm{km/s}$ and outer velocity dispersion
$\sigma\sim400.0 \rm{km/s}$. We note that the scatter in velocity
dispersion profile due simply to projection is quite substantial, with
{\it rms} variations of 10 to 15\%. Given that both the structure and
the kinematics of the simulated BCGs seem to agree well with
observation, it is interesting to investigate in greater detail how
multiple mergers shape the distributions of dark and stellar matter in
the inner regions of galaxy clusters.

\begin{figure*}
\includegraphics[width=0.4\textwidth,trim=0mm 0mm 0mm 0mm,clip]{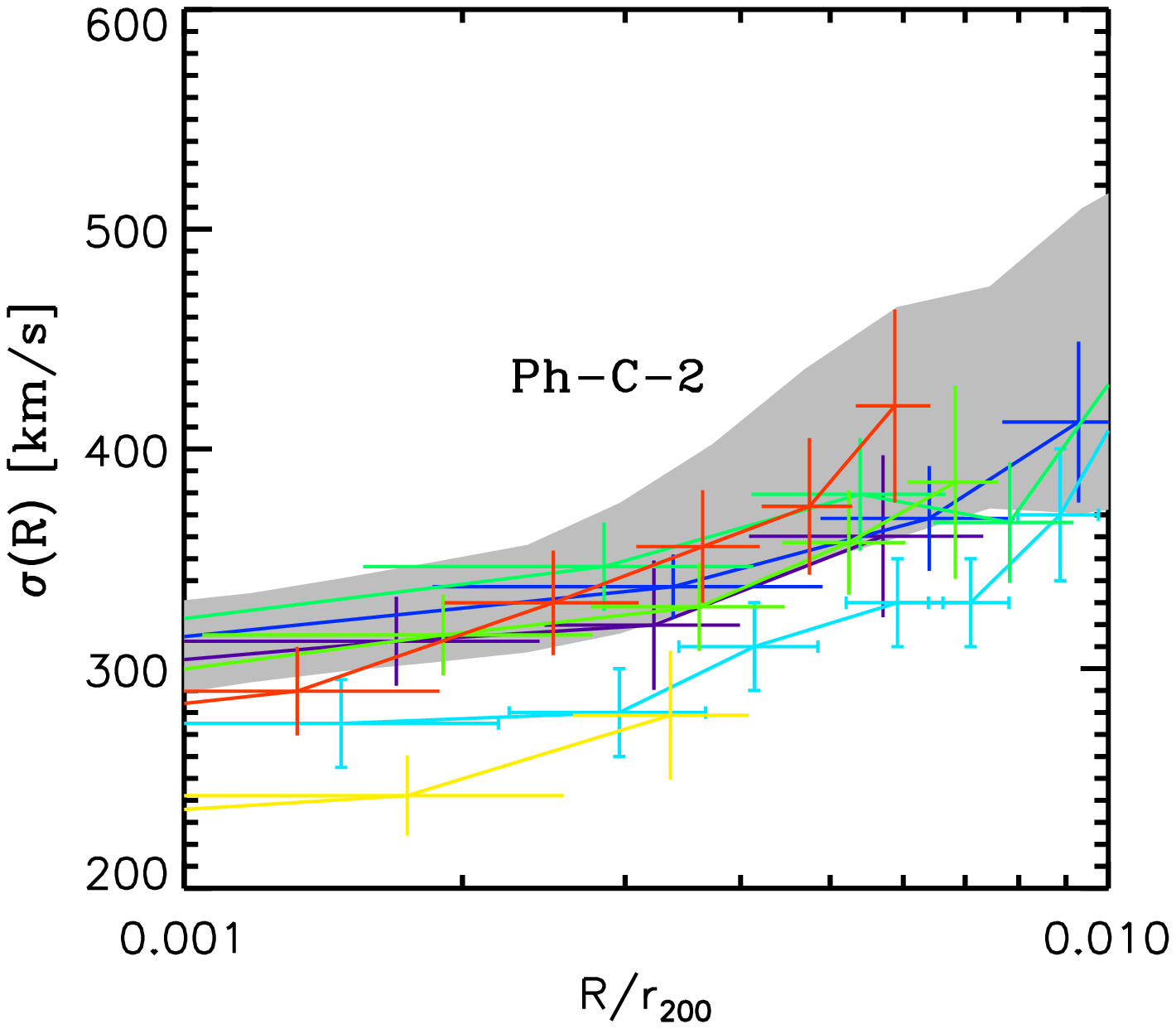}
\includegraphics[width=0.4\textwidth,trim=0mm 0mm 0mm 0mm,clip]{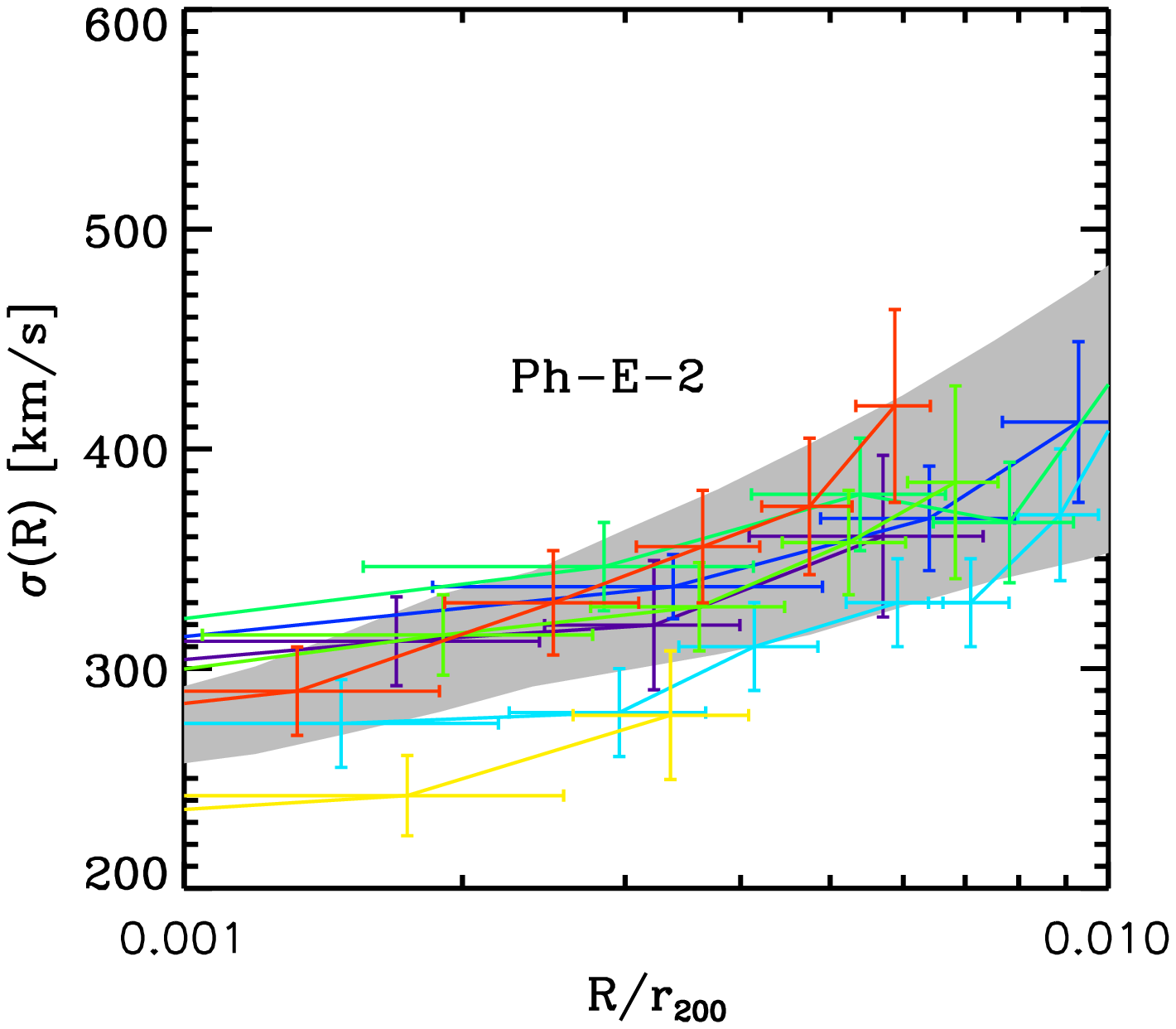}
\caption{Projected stellar velocity dispersion as a function of projected
  radius (normalised by $r_{\rm{200}}$) computed for 100 random
  orientations of our two $z=0$ BCGs. The grey area shows the $1\sigma$ scatter in
  velocity dispersion due to projection effects. Overplotted are
  the velocity dispersions measured by Newman et al. (2013a) for seven of their
  clusters.}
\end{figure*}

\section{Mergers, Mixing \& Dark Matter Heating}
\newpage

\begin{figure*}
\includegraphics[width=0.9\textwidth,trim=0mm 0mm 0mm 0mm,clip]{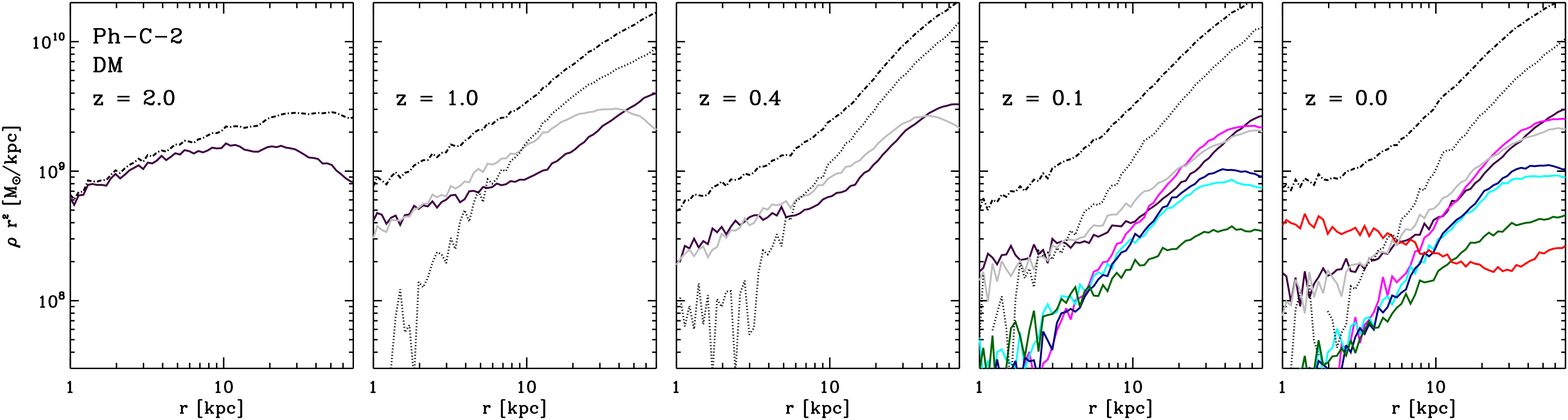}
\includegraphics[width=0.9\textwidth,trim=0mm 0mm 0mm 0mm,clip]{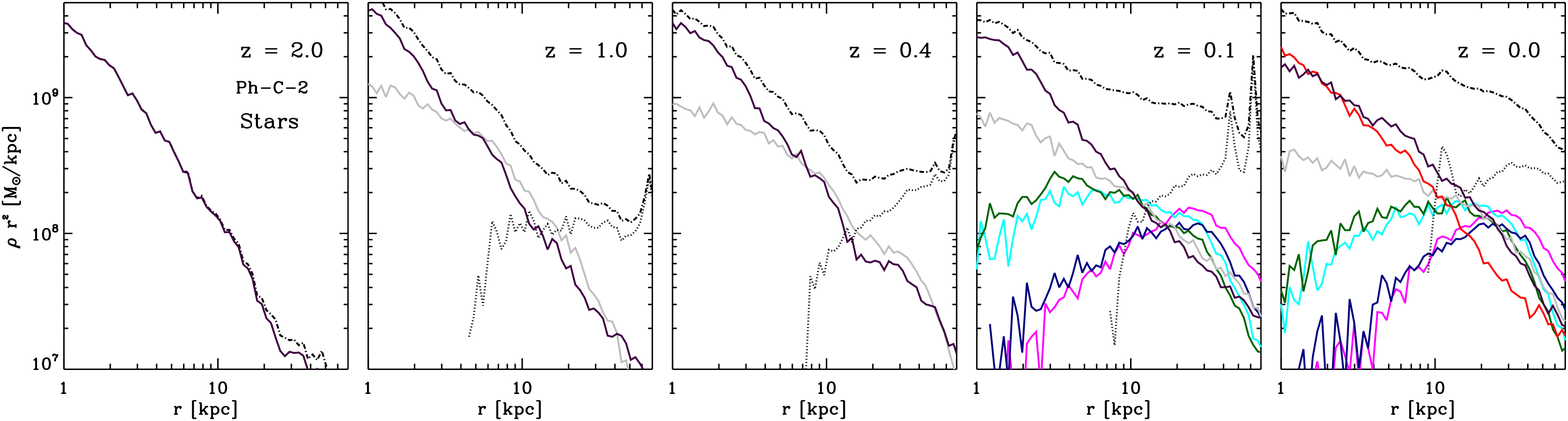}
\includegraphics[width=0.9\textwidth,trim=0mm 0mm 0mm 0mm,clip]{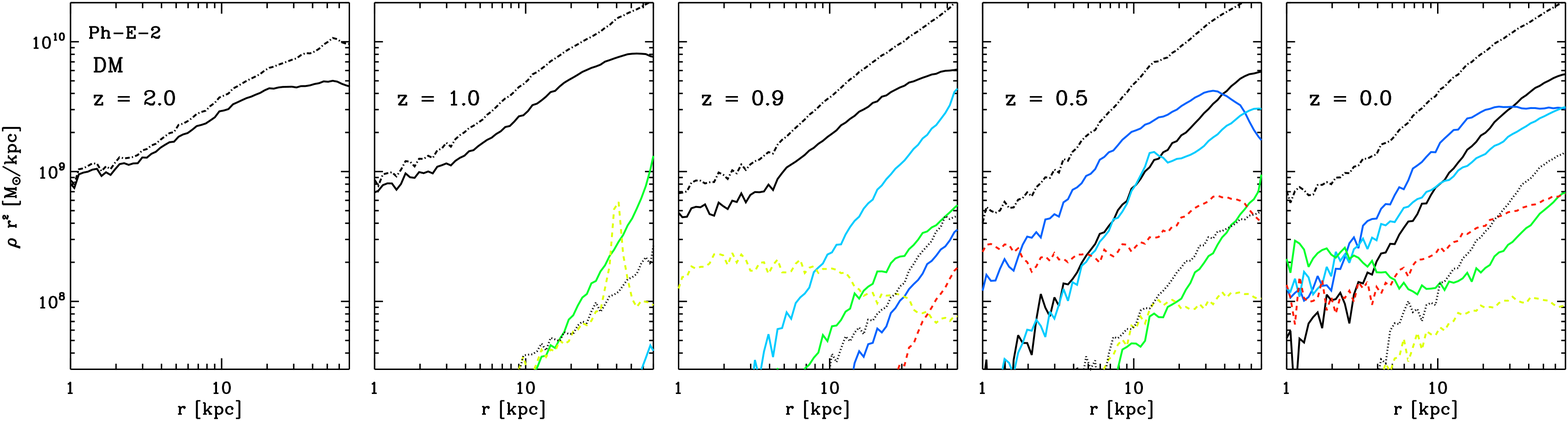}
\includegraphics[width=0.9\textwidth,trim=0mm 0mm 0mm 0mm,clip]{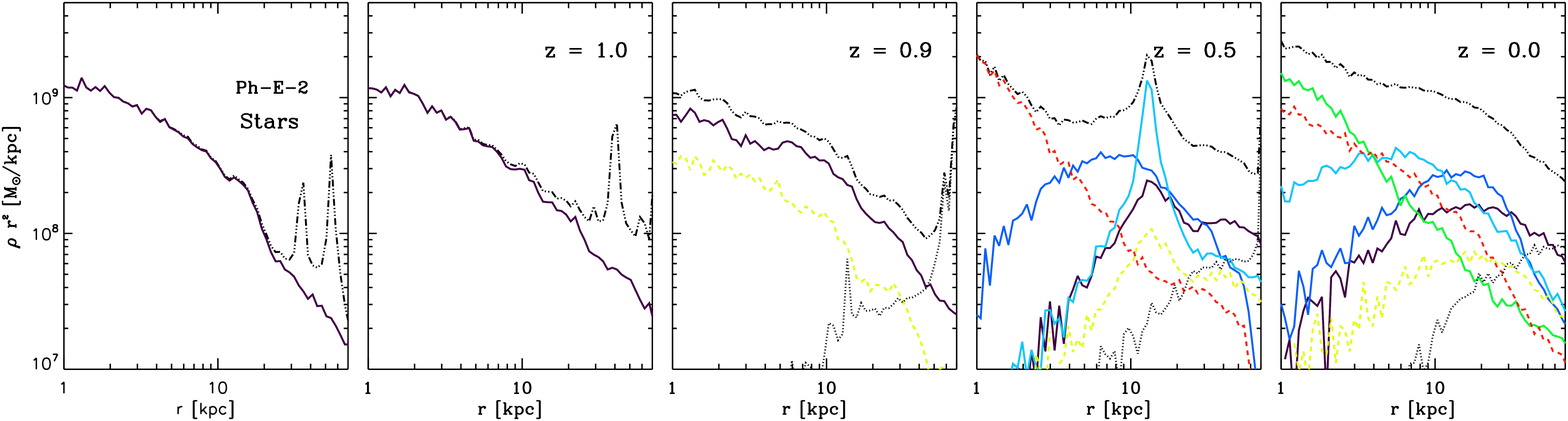}
\caption{$\rho r^2$ profiles as a function of radius for the dark
  matter and stellar components of our two clusters for particles with
  energy $E<E_{0}$ at $z=0$. The top two rows are for Ph-C-2, while
  the bottom two rows are for Ph-E-2. In each pair, the upper row
  gives dark matter profiles while the lower row gives stellar
  profiles. A thick black solid line in each panel designates the
  component associated with the most massive progenitor at
  $z=2$. Colored lines are for particles belonging to other $z=2$
  progenitors. Central and satellite objects at $z=2$ are
  distinguished by solid and dashed lines, respectively. Only profiles
  associated with $z=2$ progenitors which have brought more than a 100
  particles within 5 kpc are plotted to avoid over-crowding. Black
  dotted lines are profiles for all the material which is not part of
  any such progenitor. Combined profiles for {\it all} the stars and
  for {\it all} the dark matter at each time are plotted as
  dashed-dotted lines.}
\end{figure*}

\begin{figure*}
\includegraphics[width=0.4\textwidth,trim=0mm 0mm 0mm 0mm,clip]{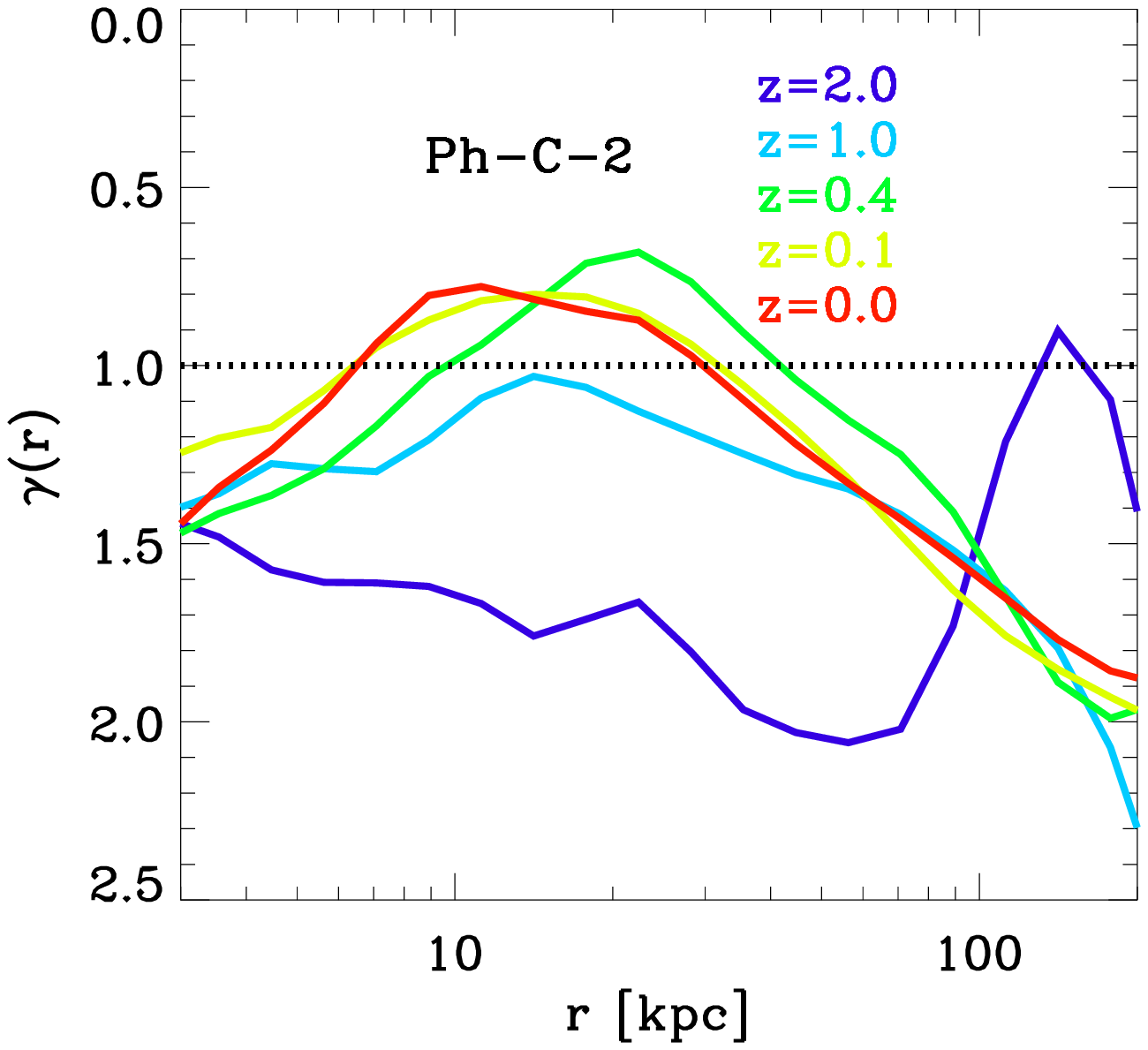}
\includegraphics[width=0.4\textwidth,trim=0mm 0mm 0mm 0mm,clip]{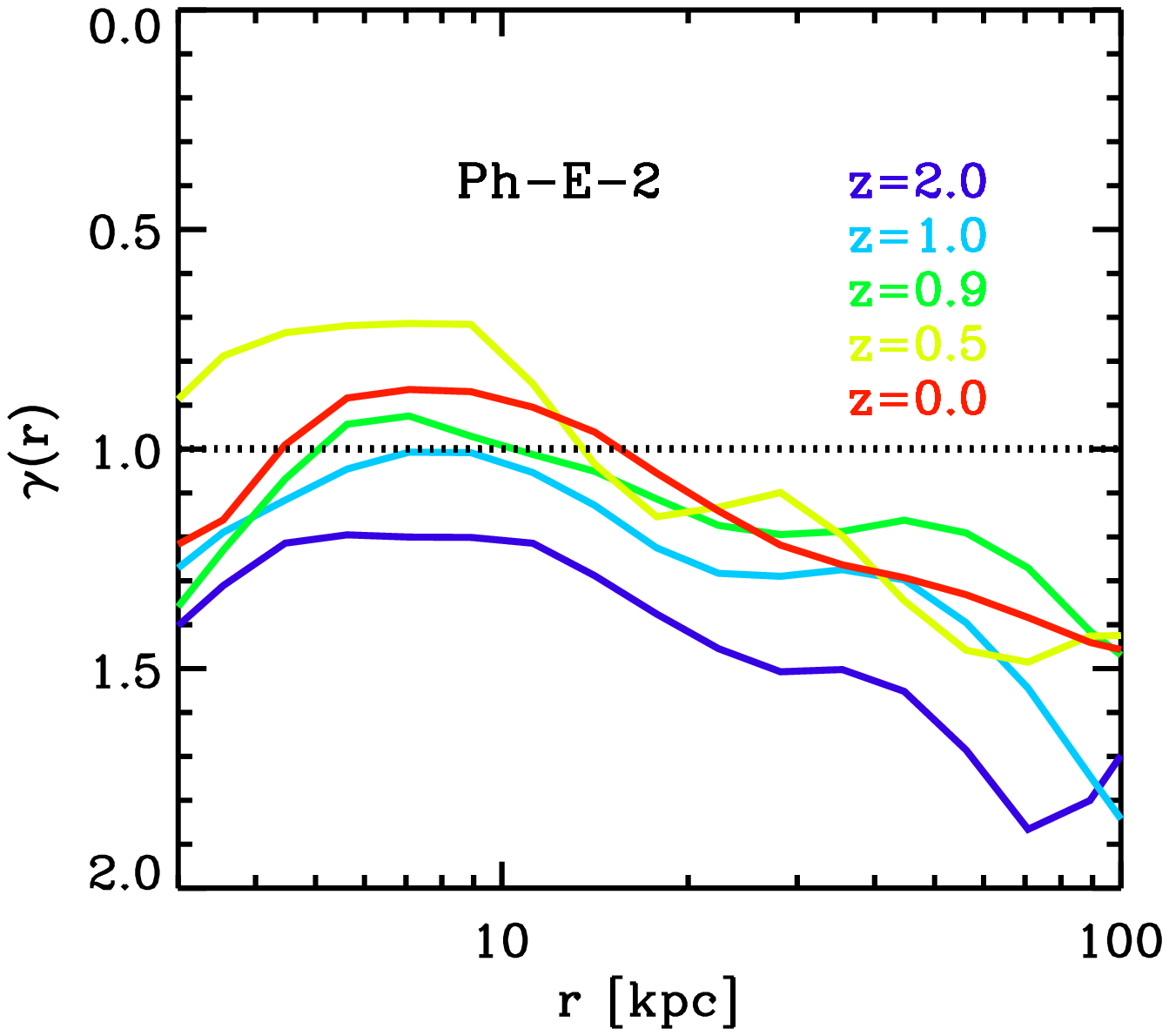}
\caption{Evolution of the slope of the dark matter profile as a
  function of radius for our two clusters. The times correspond to
  those of Figure 9. {\it Left}: Ph-C-2: A first merger before $z=1$
  has already brought the slope to quite shallow values and a later
  one below $z=0.4$ establishes a shallow dark matter cusp which is
  preserved though other mergers and extends with $\gamma<1$ from 6 to
  30 kpc at $z=0$. {\it Right}: Ph-E-2: A similar behaviour is seen
  where the slope of the dark matter gets shallower after each
  merger. Close to a merger this is more extreme (yellow line) but the
  system always relaxes to a shallower dark matter slope which by
  $z=0$ has $\gamma<1$ from 4 to $15 \, \rm{kpc}$.}
\end{figure*}

In this section, we investigate in more detail how repeated mergers
cause dark matter profiles to get shallower at small radius and total
mass profiles to converge to NFW-like form.  We proceed by selecting
particles which at $z=0$ have binding energy $E<E_{0}$, placing them
at the centre of the final galaxy cluster. $E_{0}$ is chosen so that
most of the particles with $r<100 \, \rm{kpc}$ are included. We then
track these particles back to $z=2$ and separate them according to the
progenitor objects to which they belong. We consider the most massive
of these to be the ``main'' progenitor of the final BCG. We then
follow these sets of particles forwards in time studying how they
contribute to the build up of the stellar and dark matter
distributions of the final BCG. In Figure 11 we show images of the
distribution of the stellar particles selected in this way at a series
of redshifts which we analyse in more detail below. Colored stars in
this plot indicate individual progenitors using the same colour scheme
as in the profiles of Figure 9, which we discuss next. Clearly, BCG
progenitors can be widely separated at $z=2$.

Figure 9 itself takes the form of a time series of $\rho r^{2}$
density profiles, picking times which are particularly illustrative of
the way in which mergers induce changes in structure. In these plots,
thick black lines indicate profiles from material which was part of
the main `$z=2$ progenitor while colored lines are for material from
other progenitors. The colours corespond between times and solid and
dashed lines indicate objects which at $z=2$ were central and
satellite galaxies, respectively. The black dash-dotted lines show
profiles for all the stellar/dark matter particles without separation
by binding energy or origin. Dotted lines show profiles for material
from objects which contribute fewer than 100 particles at $r<5 \,
\rm{kpc}$ (individual profiles are only shown for objects where this
is not the case). Because of the complexity in merging history and
progenitor properties (stellar mass and size) we will discuss the
plots for each resimulation individually.

\subsection{Ph-C-2: dark matter heating with no re-shuffling}

Between $z=2$ and $z=1$ the main progenitor of this cluster interacts
with a number of galaxies which substantially modify the shape of its
dark matter profile, mixing it with material from larger radii and
from other objects.  By $z=1$, one large galaxy has merged to the
centre and the system has relaxed to produce a dark matter density
profile which is already substantially shallower than initially
(compare the dash-dotted curves in the top left two panels of Figure 9
and see also Figure 10). The merging galaxy has brought in a
significant amount of dark matter which contributes at least as much
to the profile as the main progenitor at all radii (compare grey and
black curves in the second and third panels of the top row of Figure
9). The stars from the merging galaxy only dominate the merger product
beyond about 5~kpc, however. A number of other significant mergers
occur after $z=0.4$, note it is only after $z=0.1$ that another merger
occurs that effects the innermost regions of the final BCG.  Dark
matter from this last merging galaxy dominates the final profile
within a few kpc, and its stars contribute equally to those of the
main progenitor within about 10 kpc (compare the red and black curves
in the $z=0$ panels). Stars contributed by other progenitors are only
significant at larger radii.  The shallow dark matter profile
established by $z=0.4$ is counteracted slightly at the very centre by
this last merger but is maintained at all larger radii.  In this
cluster it is notable that stars from the main progenitor dominate the
inner regions of the BCG at all times except the last one, and even
then they contribute substantially down to the smallest radii.

\subsection{Ph-E-2: dark matter heating with re-shuffling}

The inner regions of this cluster show a very different assembly
history than Ph-C-2. Between $z=2$ and $z=1$, there is very little
merging activity. Essentially no stars are added to the central galaxy
and the small amount of additional dark matter comes from mixing
processes and from a few objects which were disrupted at relatively
large radius. Nevertheless, many satellites are orbiting the central
galaxy at somewhat larger radii, and this causes some mixing and
heating which causes both the stellar and dark matter profiles of the
central galaxy to become somewhat shallower.  Between $z=1$ and
$z=0.9$ one of these satellites merges with the central galaxy (see
the yellow curves in the $z=0.9$ panels -- the satellite is visible as
a spike at $\sim 45 \rm{kpc}$ in the $z=1$ panels) inducing a further small
decrease in the slopes of the stellar and dark matter profiles.
Shortly thereafter, another galaxy (indicated by light blue curves)
merges with the BCG and for a short period contributes the dominant
stellar population near its centre.  By $z=0.5$, however, two
additional galaxies (indicated by dark blue and red curves) have
merged together, and have usurped the old BCG to become the bottom of
the cluster potential well. The old BCG with its multiple components
can be seen as a concentration at $r\sim 10$ kpc from this new centre
which is disrupted into the envelope of the system at later times (see
the $z=0$ panel). The two progenitors which dominate the BCG at
$z=0.5$ contribute almost equally to its dark matter content in the
inner few kpc, but the stars in this region are contributed almost
entirely by one of the two (indicated by red curves) which had the
more compact initial galaxy. Finally, by $z=0$ yet another galaxy
(indicated by green curves) has merged into the BCG and dominates
its stellar and dark matter content in the central regions.

The underlying reason for this constant reshuffling is that mergers
tend to reduced the mass density in the inner regions of the BCG, and
as a result new inspiralling satellites can make it all the way into
the central regions without being disrupted provided they are
initially relatively compact. Thus the ``older'' BCG material is
continually pushed outwards as it is heated by the incoming objects
which are replacing it. This competition leads to the inner regions of
the final object being preferentially composed of stars that came from
compact progenitors which were added relatively late.

\subsection{The evolution of density profile slopes}

The effects of merging activity on the dark matter profile slope in
the inner regions of our two clusters can be seen more easily in Figure
10, where this slope is plotted against radius for the specific redshifts
shown in Figure 9.  These curves thus show the logarithmic derivative of
the dash-dotted curves in the dark matter panels of Figure 9. A horizontal
dotted line at $\gamma=1$ indicates the asymptotic inner slope of an NFW
profile.

In our initial conditions $\gamma$ is substantially larger than one at
all radii and in both simulations, reflecting the steepening of the
dark matter profile caused by contraction in response to the
gravitational effects of the added stars. By $z=1$ the inner dark
matter profiles have become substantially shallower, however, reaching
$\gamma=1$ in both cases, and this trend continues so that $\gamma<1$
over substantial regions by $z=0$, $6~{\rm kpc} < r < 30~{\rm kpc}$
for Ph-C-2 and $5~{\rm kpc} < r < 15~{\rm kpc}$ for Ph-E-2.

\begin{figure*}
\includegraphics[width=0.8\textwidth,height=0.4\textheight, trim=0mm 0mm 0mm 0mm,clip]{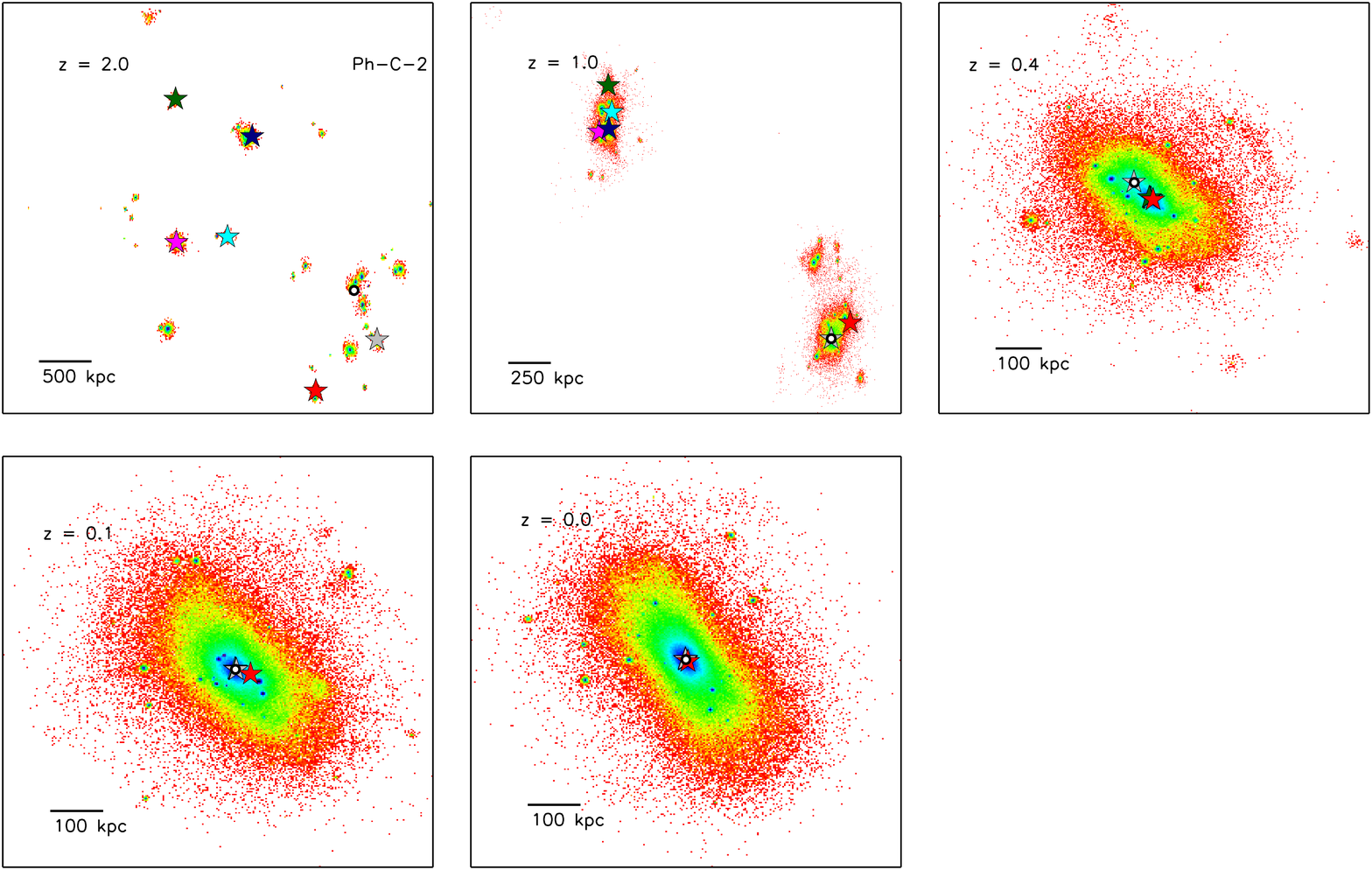}
\includegraphics[width=0.8\textwidth,height=0.4\textheight, trim=0mm 0mm 0mm 0mm,clip]{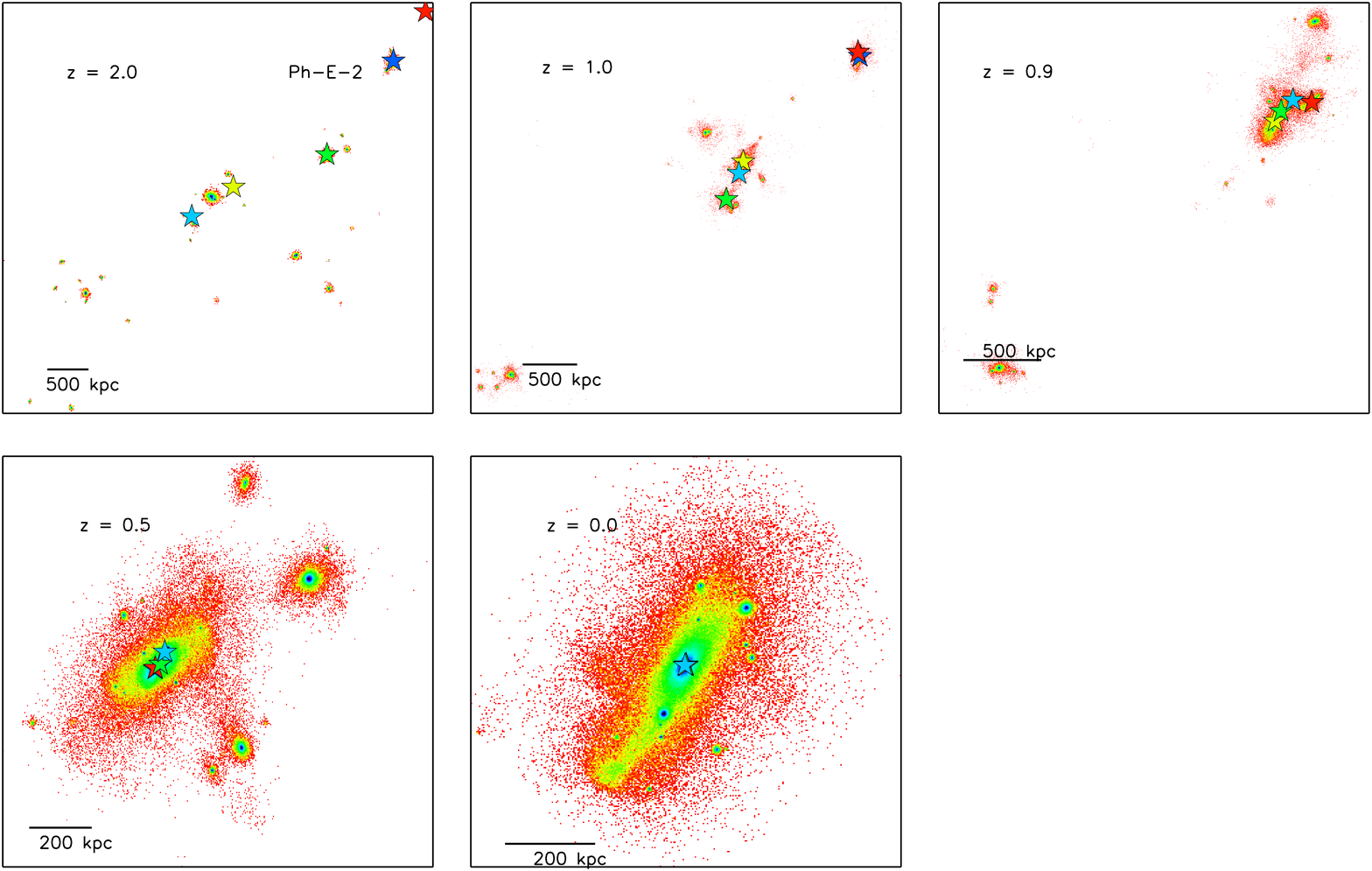}
\caption{Distribution at a series of redshifts of stellar particles
  selected to have $E<E_{0}$ at $z=0$. Only galaxies contributing more
  than 100 particles to the central $5~kpc$ of the final BCG are
  shown. The top panels are for Ph-C-2 and the lower ones for
  Ph-E-2. Stars identify individual progenitors and are colour-coded
  to correspond to the curves in Figure 9.}
\end{figure*}

\subsection{Dark matter fractions}

Athough we have simulated only two clusters, it is clear that allowing
initial galaxies to have a range of structural properties at given
mass combines with the stochasticity of merging histories to produce
very different amounts of mixing and hence a wide range of dark matter
fractions in the final BCGs. We study this as function of radius and
redshift in Figure 12. There are several ways of looking at the dark
matter fractions. One may be interested in their variation over a
given a radial range of the galaxy, which is determined by how mergers
affect the relative amounts of stars and dark matter within a fixed
physical scale. Alternatively, it may be more helpful to consider the
evolution of the dark matter fractions within a region which encloses
a given fraction of the BCG's stars, for example, within its
half-light radius. Such variations are directly related to mixing
during the growth in size and mass of the galaxy. Both types of
variation can be seen in Figure 12 for the BCG's in our two
simulations.

For Ph-C-2 the dark matter fraction as a function of radius evolves
relatively modestly, decreasing with time at all but the smallest
radii. However, the dark matter fraction within the half-light radius
increases substantially as the system evolves, from 0.1 at $z=2$ to
0.8 at $z=0$ (indicated by the values at which the black and red
arrows intersect the mass fraction curves at $z=2$ and $z=0$
respectively). In contrast, the evolution in Ph-E-2 is stronger at
fixed radius, but much milder within the half-light radius, increasing
from $f_{DM}=0.65$ to $f_{DM}=0.75$ over the same redshift range. This
illustrates that dissipationless mergers {\it can} alter the dark
matter fraction within BCGs, but the amount varies greatly from system
to system because of the diversity of structural properties in the
progenitor galaxies. Indeed, while the dark matter fraction in the
innermost regions of Ph-C-2 increases with time, that in Ph-E-2
decreases substantially from $z=2$ to $z=2$ because at late times
these regions are dominated by material from a small object that
merged with the BCG at $z\sim 0.8$ and had much higher stellar density
than the main BCG progenitor (compare with Figure 9). Interestingly,
despite the very different merger histories of our two BCG's, the
final dark matter fraction within the half-light radius agrees well
for both with estimates from \cite{Newman2013b} who find $f_{DM}\sim
0.8$ within the effective radii of their observed BCG sample.

\begin{figure*}
\includegraphics[width=0.4\textwidth, trim=0mm 0mm 0mm 0mm,clip]{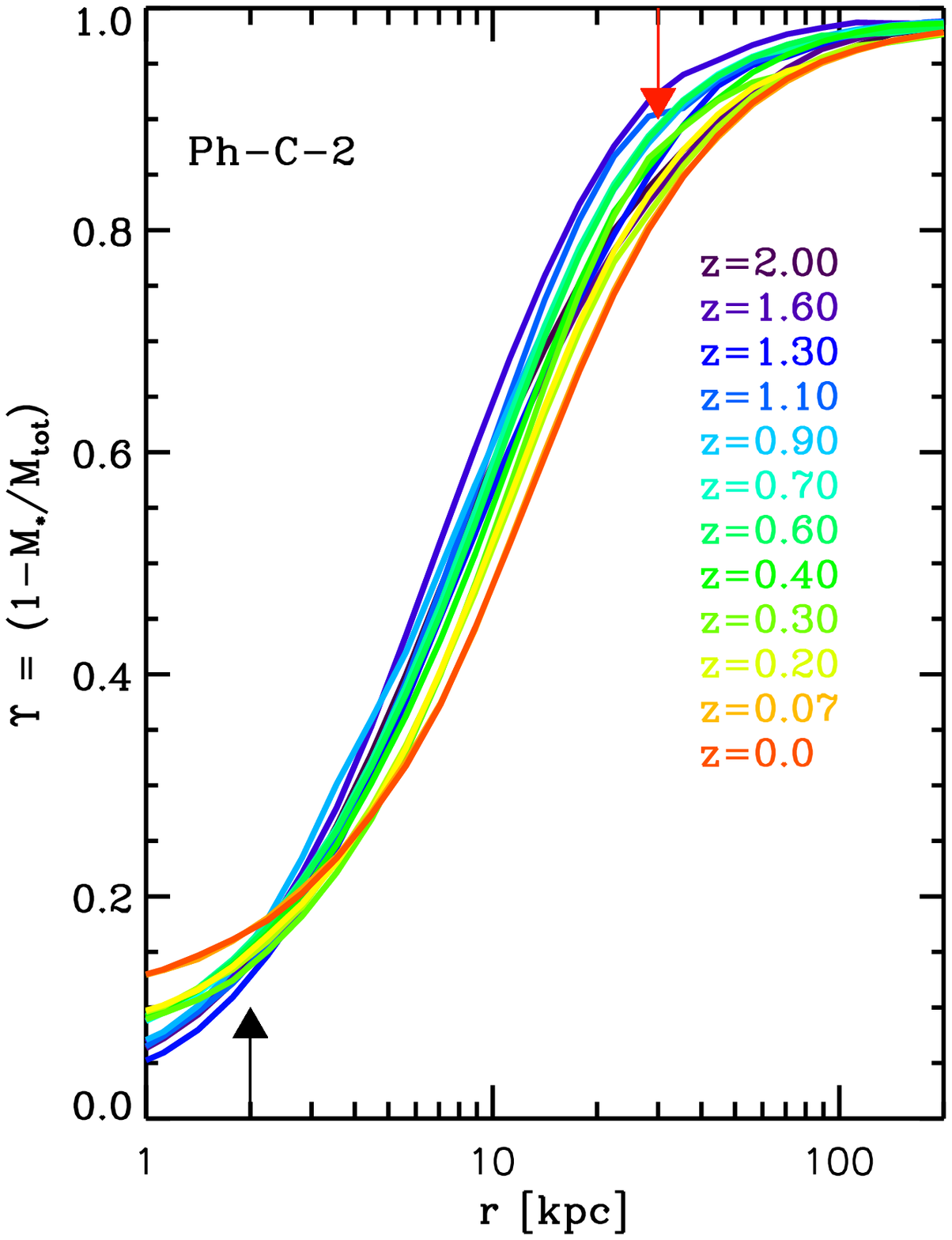}
\includegraphics[width=0.4\textwidth, trim=0mm 0mm 0mm 0mm,clip]{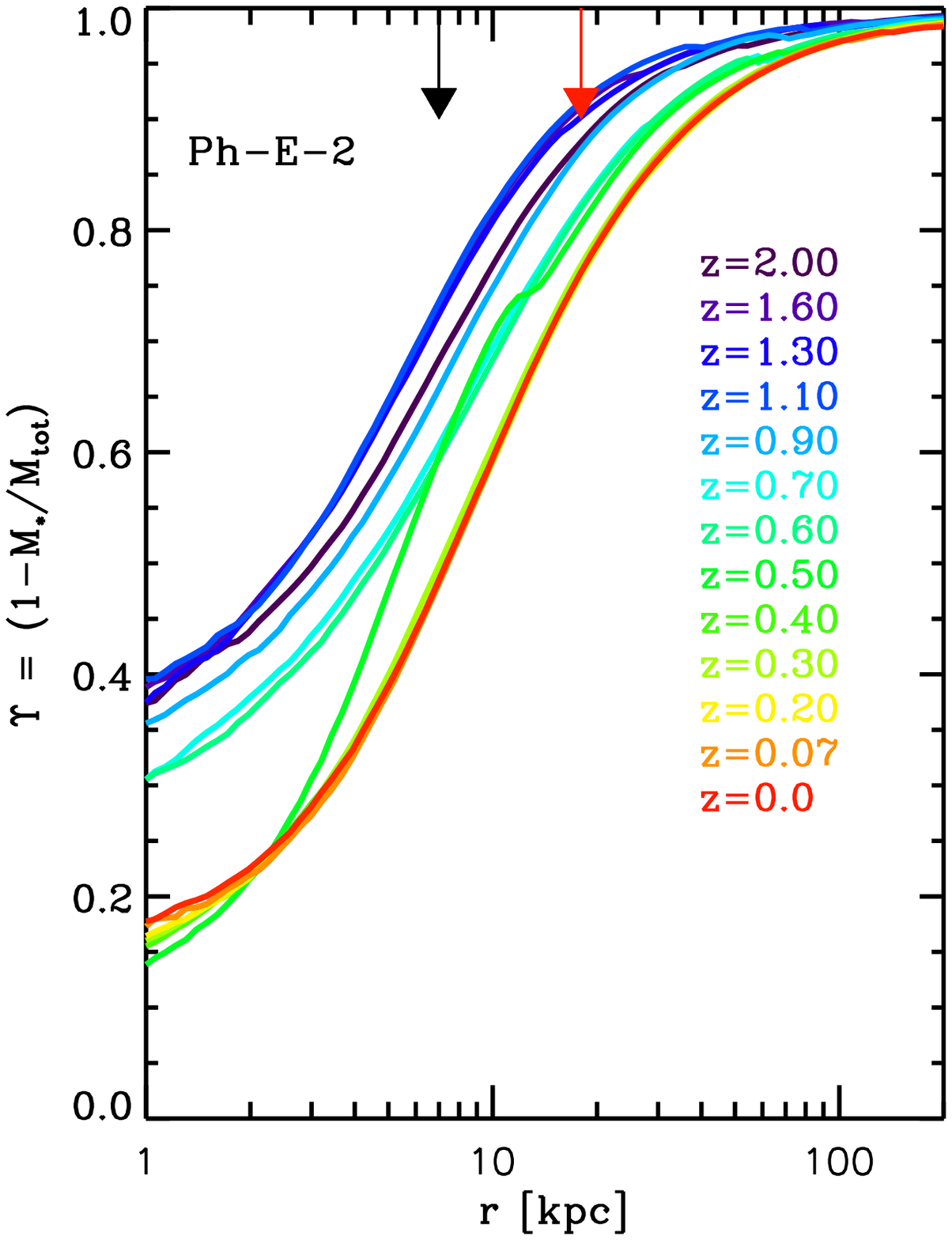}
\caption{Dark matter fraction as a function of radius for different
  redshifts (coloured lines). The black arrows indicate the half-light
  radii of the BCGs' most massive progenitors at $z=2$ and the red
  arrows those of the final BCG at $z=0$. {\it Left}: Dark matter
  fraction for Ph-C-2. The dark matter fraction increase is
  substantial: the most massive progenitor at $z=2$ had a dark matter
  fraction of 0.1 (shown where the black arrow intersects the dark
  blue solid line) within its half-light radius which increased to 0.8
  by $z=0$ (where the red arrow intersects the orange solid line).
  {\it Right}: Dark matter fraction for Ph-E-2. The dark matter
  fraction within the half-light radius of this galaxy increases
  mildly from 0.65 at $z=0$ to 0.75 by $z=0$. Note that the dark
  matter fraction decreases substantially in the innermost region of
  this galaxy (by a factor of 2 within $1 \, \rm{kpc}$ due to a merger
  with a satellite with much higher central stellar density. }
\end{figure*}

\section{The effects of supermassive black holes}

Given the substantial number of mergers occurring during the formation
of BCGs in $\Lambda$CDM, it is worthwhile to consider how the presence
of supermassive black holes may affect the central distribution of
dark and stellar matter. While the experiments of this paper do not
specifically include supermassive black holes, we can still use
observed relations between the mass of central black holes and that of
their host galaxy to estimate the potential importance of black hole
mergers in re-distributing matter in the inner regions of
BCGs. Although quite rough, such estimates are useful to give an idea
of the size of the regions which may be affected.

The BCG in Ph-E has experienced five mergers which have brought
significant stellar material to the centre of the galaxy. From the
initial stellar masses of the progenitor galaxies, we can estimate the
masses of the black holes they host. We do this using the relation
published by \cite{Bennert2011}: $\log(M_{\bullet})=\alpha(\log(M_{*})-10)
+ \beta\log(1+z) +\gamma + \sigma +8 $, with $\alpha=1.09$,
$\beta=1.96$, $\gamma=-0.48$, $\sigma=0.36$.  When two galaxies merge,
their central black holes are assumed to merge also, and in its later
stages the in-spiral of the binary black hole deposits energy in the
inner regions of the stellar merger remnant, giving rise to an
apparent ``mass deficit'' relative to the structure expected in the
absence of the black holes. \cite{Merritt2006} used N-body simulations
to study this process and to relate the size of this stellar mass
deficit to the masses of the merging black holes.  He found
$M_{\rm{def}}\sim 0.5 M_{12}$, with $M_{12}=M_1+M_2$ the total mass of
the binary: suprisingly this result depended only weakly on the mass
ratio $q=M_{1}/M{2}$ of the black holes or on the inner structure of
the stelar density distribution. Assuming that after each of a series
of $N$ mergers with $M_{2} \ll M_{\bullet}$ the black holes coalesce
to a new central object, he estimated the cumulative mass deficit as
$M_{\rm{def}}=0.5 N M_{\bullet}$, where $M_{\bullet}$ is the final
mass of the black hole.

We apply this to Ph-E-2 by identifying $M_{\bullet}$ as the total mass
of all the black holes assigned to the $z=2$ progenitor galaxies which
merge to the centre of the BCG. We then compare this to the enclosed
total mass within 3, 2, 1 $\mathrm{kpc}$. If all black-holes were to
merge, this final black-hole mass would be $M_{BH}=6.0\times10^{10}
\mathrm{M_{\odot}}$.  We will assume that after the first merger, a
binary black hole is formed with mass $M_{12}=M_{1}+M_{2}$. In this
case $M_{12}=2\times 10^{10} \mathrm{M_{\odot}}$. We also note that
the black holes involved in subsequent mergers have $M<<M12$. So given
that $N=4$ further mergers occurred at the centre of the BCG, our
estimated deficit mass would be $M_{def}\sim 0.5 N M_{\bullet} \sim
12.0 \times 10^{10} \mathrm{M_{\odot}}$. This could easily account for
a core radius of at least $3 \mathrm{kpc}$, given that the simulated
merger remnant has $M(r<3\mathrm{kpc})=1.0\times 10^{11}
\mathrm{M_{\odot}}$. Such large cores are visible in some BCGs
\citep{Postman2012} and, based on this experiment, can be well
accounted for in $\Lambda$CDM.

Proceeding in a similar way for Ph-C-2, we identify six mergers. Given
the $z=2$ progenitors in this system, these would lead to a final
black hole mass of $M_{\bullet}\sim7.4\times10^{10}
\mathrm{M_{\odot}}$ and a corresponding mass deficit of $M_{def}\sim
0.5 N M_{\bullet}\sim 1.8 \times 10^{11} \mathrm{M_{\odot}}$. Again,
this is enough to explain core sizes of $3 \mathrm{kpc}$ or more,
given that $M(r<4\mathrm{kpc})=2.0\times10^{11} \mathrm{M_{\odot}}$ in
this BCG.  For both clusters we show the radii where mass deficits may
affect the distribution of matter in Figures 6 and 7.

We stress however, that these estimates assume that all the black
holes coalesce in a series of binary mergers. This may not be the case
since triple systems could form and lead to sling-shot ejection of one
or all black holes \cite{Mikkola1990}. This would result in smaller
values of $M_{\bullet}$ but larger values of $M_{def}$ because
multiple black hole systems are more efficient than a binary to move
stars around \citep{Merritt2004,Boylan-Kolchin2004}. Gas may also be
expected in some of the progenitor galaxies (given constraints on the
fraction of quiescent galaxies at $z=2$ \citep{Muzzin2013}) hastening
the coalescence of some binaries, but also promoting the formation of
stars and hence smaller mass deficits. Of course, the $z=2$
star-forming galaxies may already have lost their gas by the time of
the merger, which would eliminate such dissipational effects.

These back-of-the-envelope calculations based on our experiments
suggest that black holes may play an important role in the
re-distribution of matter within the inner $\sim 3~\mathrm{kpc}$ of
BCGs, reducing the densitites of both stars and dark matter in these
regions.  This would improve the agreement of the experiments of this
paper with the observational data of \citet{Newman2013a} and so would
be an interesting topic for further investigation.

\section{Discussion}

Earlier work on this topic has argued that dark matter cusps in the
inner regions of clusters will be weakened by energy input from
dynamical friction as galaxies merge into the central object, but has
failed to be fully convincing through overly idealised treatments and
the lack of full gravitational consistency \citep{El-Zant2004,
  Nipoti2004, Laporte2012}.  Our current experiments do not suffer
from such strong approximations and have been designed to test the
importance of dissipationless mixing processes at the centre of galaxy
clusters in a realistic cosmological context. Our set-up assumes that
at late times the assembly of the inner regions of galaxy clusters is
entirely dominated by collisionless merger processes. This is
motivated by theoretical work on the formation and evolution of BCGs
\citep{Delucia2007, Laporte2013} which appears consistent with
observed evolution \citep{Gonzalez2005, Stott2011, Lidman2012}. Such
an assumption makes the problem well-posed and addressable through
N-body simulations: present-day BCGs should be formed from the
population of galaxies observed at $z=2$ by dynamical processes acting
in the concordance $\Lambda$CDM cosmology. The consistency of our
results with the BCG structure which \cite{Newman2013a} infer for a
massive cluster sample including both cool-core and non-cool-core
clusters suggests that dissipationless assembly may indeed dominate
the late-time growth in all clusters. We note in passing that AGN
feedback may also act to produce shallow dark matter cusps
\citep[e.g.][]{Martizzi2011}, but current results are not
quantitatively convincing because proper hydrodynamical modelling is
not possible at full resolution and predictions are highly
resolution-dependent \citep{Choi2014}. Feedback effects on the
dynamics of stars and dark matter at the centres of galaxies are
undoubtedly an interesting area for further investigations.

Our initial conditions were created through impulsively adding stars
at the centres of $z=2$ dark matter haloes. This causes a transient
re-virialisation phase and a compression of the inner regions of the
final equilibrium halo. The extent of such compression for realistic
galaxy formation models has long been debated \citep[see, for example,
  the discussion in][]{Tissera2010}. It is clear, however, that the
inner total mass profiles of our initial galaxies at $z=2$ lie well
above the NFW expectations in a dark-matter-only simulation, both
because of this compression and because of the added stars.  This
makes it all the more remarkable that multiple dissipationless mergers
can reshape the central mass distribution in the cores of galaxy
clusters so that the total mass ends up with an NFW profile and the
dark matter distribution has a shallower central cusp.

Below $r \sim 4 \rm{kpc}$ our simulations are are unable to predict
the $z=0$ density structure of matter realistically because we do not
include the effects of supermassive central black holes.  Currently,
this problem is difficult to address in cosmological simulations
because of the very high spatial and mass resolution which it
requires. However, we have estimated how much mass (both dark matter
and stars) may be pushed out of the inner regions by black hole
mergers, based on the study of \cite{Merritt2006}. Such effects are
predicted to impose an apparent core with radius $r\sim 3
\mathrm{kpc}$ and to reduce mass densities even at somwhat larger
radii. This has implications for the dark matter annihilation signal
predicted from cluster centre \cite{Gao2012}. When black hole effects
are included, the number of mergers we see in our simulations is
sufficient to explain even the largest stellar cores observed in some
BCGs \cite{Postman2012}.

Our experiments do not fully validate the attractor hypothesis of
\cite{Loeb2003} because the innermost regions of our final clusters
(which are entirely dominated by stars from one or two galaxies) do
not follow the profiles found in the dark-matter-only runs.  The
profiles are, however, similar in the two cases in the regions where
there is significant mixing between dark matter and stars. Observed
galaxy structure at high redshift ($z\sim2$) covers a wide range of
sizes at given stellar mass (an order of magnitude roughly), but
clearly a much broader range (or a very different one) is possible in
principle. Thus the near ``universal'' profile we recover may be
related to the particular way in which stellar structure scales with
halo mass. It is thus hard to asses whether an attractor exists under
more general conditions. An idealised experiment of the sort carried
out by \cite{Arad2005}, but considering a family of cuspy models,
could shed light on this question. Nevertheless, it is interesting to
see that the profiles observed by \cite{Newman2013a,Newman2013b} are
very close to those expected in a $\Lambda$CDM universe where
high-redshift galaxies have the structure observed at $z=2$ and the
subsequent assembly of BCGs occurs purely through the dissipationless
mergers predicted by the $\Lambda$CDM model.

\section{Conclusion}

We have presented a series of collisionless N-body resimulations of
the growth of rich galaxy clusters between $z=2$ and $z=0$. We show
that dissipationless mergers are expected to produce shallow dark
matter cusps consistent with the recent detailed study of
\cite{Newman2013b}. The profiles in our simulated clusters drop below
the NFW expectation at $r/r_{200}\sim0.1-0.2$, as observed. These
radii are close to the half-light radii of the BCGs. At smaller radii,
the dark matter profiles in our simulations including stars are
shallower by $\Delta\gamma\sim 0.3-0.4$ than in their dark-matter-only
counterparts, despite the fact that these profiles were actually
steeper in the initial $z=2$ galaxies. Multiple dissipationless
mergers have a strong effect on the dark matter and stellar
distributions and lead naturally to the striking result of
\cite{Newman2013b} that the {\it total} mass profile (stars + dark
matter) in the cluster core has near-NFW form, and is close to that
found in dark-matter-only simulations. Our agreement with observation
breaks down in the inner few kpc, but we argue that mergers between
the supermassive black holes found in the centres of all massive
galaxies are expected to reshape the stellar (and dark matter)
distribution of the BCG within $3-4 \, \rm{kpc}$, giving rise to the observed
cores.

An interesting aspect of our simulations is the illustration that the
apparent universality of dark matter density profiles may not be set
entirely by the physics of early virialisation but may also reflect
energy exchange during subhalo mergers. In our case, the increased
mass density near the centre of our initial objects is substantially
weakened by subsequent mergers so that the total density profile
asymptotes back to that found for evolution from unmodified (i.e.
dark-matter-only) initial conditions.  The challenge in understanding
the emergence of this universal solution at a more detailed level is
related to analysing the competing effects of dynamical friction,
which transfers energy from inspiralling clumps to the diffuse
background, and of tidal stripping of the clumps, which adds new
material to the diffuse background. In any case, our numerical
experiments suggest that the total density profiles measured by
\cite{Newman2013a} can be interpreted as validating the galactic
cannibalism scenario proposed as the formation channel of BCGs by
\cite{Ostriker1975} and \cite{White1976}.

\section*{Acknowledgments}
CL was supported in part by by the Marie Curie Initial Training
Network CosmoComp (PITN-GA-2009-238356). SW was supported in part by
ERC Advanced Grant 246797, ``Galformod''. CL is supported by a Junior Fellow award from the Simons Foundation. CL would like thank Thorsten Naab, Carlos Frenk, Julio Navarro for useful discussions during the beginning of this project and Scott Tremaine for his comments on an earlier draft.

\bibliographystyle{mn2e}
\bibliography{master2.bib}{}
\label{lastpage}
\end{document}